\documentclass[journal=jacsat,manuscript=article]{achemso}

\usepackage{achemso}
\usepackage{graphicx}
\usepackage{caption}
\usepackage{subcaption}
\usepackage{amsmath}
\usepackage{amsbsy}
\usepackage{amsthm}
\usepackage{amsfonts}
\usepackage{float}
\usepackage{sidecap}
\usepackage{mathtools}
\usepackage{braket}
\usepackage{adjustbox}

\usepackage[utf8]{inputenc}
\usepackage[T1]{fontenc}
\usepackage{graphicx}
\usepackage{longtable}
\usepackage{wrapfig}
\usepackage{rotating}
\usepackage[normalem]{ulem}
\usepackage{amsmath}
\usepackage{amssymb}
\usepackage{float}
\usepackage{colortbl}
\usepackage{capt-of}
\usepackage{hyperref}
\usepackage{tabularx}

\author{Jiaqi Yang}
\email{yang1494@purdue.edu}
\affiliation[]
{School of Materials Engineering, Purdue University, West Lafayette, IN 47907, USA}

\author{Arun Mannodi-Kanakkithodi}
\email{amannodi@purdue.edu}
\affiliation[]
{School of Materials Engineering, Purdue University, West Lafayette, IN 47907, USA}

\title{First Principles Investigation of Polymorphism in Halide Perovskites}

\begin{document}

\begin{abstract}
Halide perovskites have been extensively studied as materials of interest for optoelectronic applications. There is a major emphasis on ways to tailor the stability, defect behavior, electronic band structure, and optical absorption in halide perovskites, by changing the composition or structure. In this work, we present our contribution to this field in the form of a comprehensive computational investigation of properties as a function of the perovskite phase, different degrees of lattice strains and octahedral distortion and rotation, and the ordering of cations in perovskite alloys. We performed first principles-based density functional theory computations using multiple semi-local and non-local hybrid functionals to calculate optimized lattice parameters, energies of decomposition, electronic band gaps, and theoretical photovoltaic efficiencies. Trends and critical observations from the high-throughput dataset are discussed, especially in terms of the range of optoelectronic properties achievable while keeping the material in a  (meta)stable phase or distorted, strained, or differently ordered polymorph. All data is made openly available to the community and is currently being utilized to train state-of-the-art machine learning models for accelerated prediction and discovery, as well as to guide rational experimental discovery.
\end{abstract}

\section*{Introduction}

Halide perovskites (HaPs) are very attractive materials for a variety of electronic and optical applications, primarily owing to their massive chemical space and engineerability in terms of composition, structure, alloying, and doping \cite{HP1,HP2,HP3,HP4,MRS_Bulletin_JY_Arun}. The canonical ABX$_3$ perovskite family of materials are of great interest as absorbers in solar cells \cite{OP1,OP2}, with record efficiencies achieved to date of 25.7\% in single-junction solar cells \cite{perovs_sj_record} and very recently, 32.5\% in Si-perovskite tandem solar cell \cite{perovs_mj_record,nrel}. While these efficiencies firmly place HaPs in the fastest growing market for solar absorption, they are also well below theoretical maximum values \cite{SQ_lim1,SQ_lim2,SQ_lim3}. The sheer size of the possible HaP chemical space, including 3D and layered materials, purely inorganic and hybrid organic-inorganic perovskites (HOIPs), and complex alloys, makes it difficult to screen promising materials in a brute-force manner, but also provides massive opportunities for discovery and understanding via high-throughput computation. \\

The published literature contains several glittering examples of data-driven and experimental efforts to optimize perovskite compositions for optoelectronic performance \cite{HP_disc1,HP_disc2,HP_disc3,HP_disc4,HP_disc5}. However, comprehensive understanding of the effect of polymorphism on the properties of HaPs is still an active area of research. The cubic phase is the standard prototype structure any perovskite is simulated in, and there are established numerical metrics such as the Goldschmidt tolerance and octahedral factors that determine the stability of any ABX$_3$ compound in the cubic phase \cite{Bartel}. As shown in our recent work, such factors are important but not sufficient conditions for perovskite stability, as a thermodynamic evaluation based on first principles reveals many materials that may decompose to alternative phases despite suitable ionic radii \cite{Mannodi-HP4}. Perovskites could further adopt a series of other prototype phases such as tetragonal, orthorhombic, or hexagonal, as well as other corner-, edge-, or face-shared phases such as distorted orthorhombic and needle-like \cite{MRS_Bulletin_JY_Arun,CP1,CP2,HCP1}. Polymorphism may also manifest within the same phase, in terms of energetically favorable (or metastable) distortions or rotations in corner-shared BX$_6$ octahedra or via uni-axial or multi-axial lattice strains \cite{oct1,oct2,oct3,oct4}, as well as in terms of re-optimization of the compound in larger supercells with symmetry-breaking via small distortions \cite{Dalpian}. The result is typically the existence of multiple competing phases that may all contribute as an ensemble to experimentally measured band gaps and optical absorption, instead of a sole ground state structure determining the properties. \\

Doping and alloying are two of the most common ways to engineer the properties of perovskites; the former typically involves a heterovalent ion of a suitable size substituting an A or B cation \cite{dop1,dop2,dop3}, whereas the latter involves multiple homovalent cations or anions mixed together at A/B or X sites \cite{alloy1,alloy2,alloy3,alloy4}, respectively. The number of ways to dope HaPs or create mixed compositions are practically infinite, introducing several degrees of freedom in the HaP structure-composition-properties space \cite{MRS_Bulletin_JY_Arun}. In multiple recent studies, we comprehensively explored both B-site dopants \cite{Mannodi-HP1,Mannodi-HP2} and A/B/X-site alloying \cite{Mannodi-HP3,Mannodi-HP4} in ABX$_3$ compounds using density functional theory (DFT) computations, and examined how the stability and optoelectronic properties depend on the nature of ionic mixing. We used the special quasi-random structures (SQS) approach \cite{SQS} to simulate alloys in large supercells, as is the norm in the literature, but for any given mixed composition, the properties may depend heavily on ionic ordering as well, creating more polymorphs that must be considered. \\

\begin{figure}[h]
\begin{center}
 \includegraphics[width=1.0\textwidth]{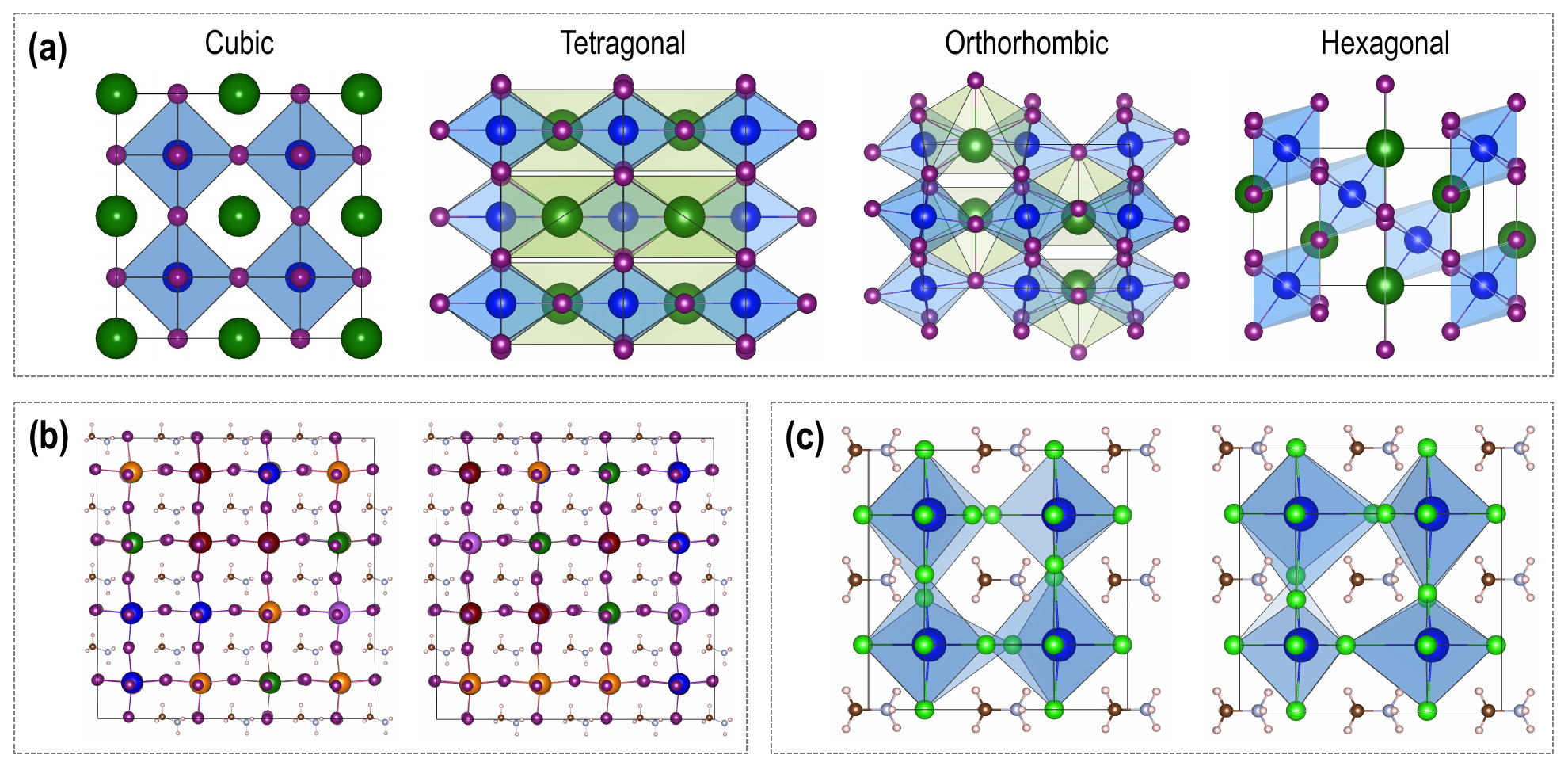}
  \caption{(a) 4 prototype ABX$_3$ HaP phases, namely cubic, tetragonal, orthorhombic, and hexagonal. (b) 4$\times$4$\times$4 cubic supercells showing a MA(Pb-Sn-Ba-Sr-Ca)I$_3$ quinary alloy with different ionic ordering. (c) Octahedral distortion in the MAPbBr$_3$ cubic lattice.} \label{fig:structs}
  \end{center}
\end{figure}

Additionally, the difficulties of brute-force experimentation within a massive structure-composition space means that high-throughput DFT (HT-DFT) computations are essential for systematically assessing trends and correlations that may guide multi-objective optimization and rational experimental synthesis and testing. DFT is extensively applied for determining lattice parameters, heat of formation or decomposition ($\Delta$H), band gap (E$_{g}$), optical absorption-derived spectroscopic limited maximum efficiency (SLME) \cite{slme1,slme2}, and defect formation energy (DFE) in perovskites, with mixed accuracy compared to experiments \cite{MRS_Bulletin_JY_Arun,T_Das}. While semi-local GGA-PBE and variants such as PBEsol (improved PBE for solids) \cite{PBEsol} and PBE-D3 (for weak dispersion interactions) \cite{PBE-D3} reproduce bulk stability and structure well, they generally under-predict E$_{g}$ compared to non-local hybrid HSE06 functional or beyond-DFT GW approximation \cite{T_Das,MKYC_gap}. For many Pb/Sn HOIPs, PBE E$_{g}$ without including spin-orbit coupling (SOC) is often as accurate as HSE E$_{g}$ with SOC included \cite{MRS_Bulletin_JY_Arun,Mannodi-HP3,Mannodi-HP4}, an effect that holds true for DFEs and corresponding defect charge transition levels (CTLs) \cite{Mannodi-HP2,Yan_defects}. HSE+SOC is more expensive but generally more accurate for electronic properties. Oftentimes, the Hartree Fock to semi-local exchange parameter $\alpha$ needs to be tuned in HSE \cite{HP_hse}, which is highly sensitive to material composition and not easy to perform over massive chemical spaces. Thus, the DFT functional itself is an added factor that determines perovskite properties: while general property-polymorph relationship trends may be reliable from semi-local functionals, more advanced theories would be necessary for quantitative estimates that could be compared with experiments. \\

Based on the above ideas, and building upon our past work, we present here a systematic investigation of the following types of polymorphism in a selected chemical space of ABX$_3$ HaPs, employing different types of PBE and HSE06 functionals within DFT:

\begin{enumerate}
    \item The effect of perovskite phase, as it changes from cubic to tetragonal to orthorhombic to hexagonal. The four phases are pictured in \textbf{Fig. \ref{fig:structs}(a)}.
    \item The effect of ionic ordering in B-site mixed compounds simulated in large supercells; two example structures for a quinary alloys are pictured in \textbf{Fig. \ref{fig:structs}(b)}.
    \item The effect of lattice strain and octahedral distortion/rotation within the cubic perovskite lattice, as shown in \textbf{Fig. \ref{fig:structs}(c)}.
\end{enumerate}

This work provides an understanding of how the above factors may positively or adversely affect the HaP stability and properties of interest for optoelectronic applications, especially single-junction solar absorption. Within a chemical space defined by A = FA, MA, or Cs (where FA and MA are organic molecules, formamidinium and methylammonium, respectively), B = Pb, Sn, Ge, Ba, Sr, and Ca, and X = I, Br, and Cl, we define easily-attainable and generalizable descriptors that encode the composition, phase, elemental properties, and ionic ordering, leading to important design rules such as the favorability of Ba-Ba clustering in increasing the bulk stability and the extent to which strain and distortions may keep the lattice stable while tuning the band gap. We also obtain important insights on the accuracy of different functionals and the inter-relationships between them. We believe the datasets and understanding obtained from this work will be crucial for guiding subsequent studies, both computational and experimental, as well as training machine learning (ML) models for accelerated prediction and screening over hundreds of thousands of possible structures and compositions. In the following sections, we describe computational details and present a series of plots and discussions unraveling the effects of different factors on the properties of interest. All data is made openly available for the benefit of the community. \\

\section*{Computational Methodology}

\subsection*{\textit{DFT Details}}

All DFT computations were performed using VASP version 6.2 \cite{vasp1,vasp2,vasp3}, employing Projector Augmented Wave (PAW) pseudopotentials. \cite{PAW1,PAW2} The Perdew, Burke, and Ernzerhof (PBE) functional within the generalized gradient approximation (GGA) \cite{vasp_pbe} as well as the Heyd-Scuseria-Ernzerhof (HSE) functional (\(\alpha\)=0.25 and \(\omega\)=0.2) are used for the exchange-correlation energy. The energy cutoff for the plane-wave basis is set to 500 eV. A Monkhorst-Pack k-point mesh of 6$\times$6$\times$6 is used for cubic unit cells and a mesh of 4$\times$4$\times$3 is used for prototypical tetragonal, orthorhombic, and hexagonal unit cells. For cubic supercell calculations, the k-point meshes are reduced to 3$\times$3$\times$3 and gamma-point only for 2$\times$2$\times$2 and 4$\times$4$\times$4 supercells, respectively. The k-point meshes are accordingly scaled down for the tetragonal, orthorhombic, and hexagonal supercells as well. Starting from the PBE-optimized structures, full geometry optimization is additionally performed using PBEsol, PBE-D3, and PBEsol-D3, by adding relevant input tags. The force convergence threshold is set to be -0.05 eV/{\AA} for all geometry optimization runs. Spin-orbit coupling (SOC) is incorporated in HSE06 calculations using the LORBIT tag and the non-collinear magnetic version of VASP 6.2 \cite{VASP_SOC}. \\

The PBE-optimized structure is used as input for calculating the optical absorption spectrum using the LOPTICS tag and setting the number of energy bands to 1000, and the approach developed by Yu et al. \cite{yu-2012-ident-poten} is then applied to determine the spectroscopic limited maximum efficiency (SLME) as a function of sample thickness. The SLME value at 5 $\mu$m thickness is taken as the theoretical photovoltaic (PV) efficiency. The band gap is computed from the PBE optimization runs and from static HSE calculations based on the PBE-optimized structure, where k-point meshes of 2$\times$2$\times$2 and 2$\times$2$\times$1 are respectively used for cubic and tetragonal/orthorhombic/hexagonal supercells. SLME values at the PBEsol, PBE-D3, PBEsol-D3, and HSE levels are determined by shifting the PBE-computed optical spectra by the difference between the PBE band gap and that computed from the corresponding functional, and recalculating the SLME based on the approach used in past work \cite{Mannodi-HP4}. We consider a series of pure and mixed-composition ABX$_3$ compounds in this work; all alloys are simulated using SQS except for the 4$\times$4$\times$4 supercell structures, where we explicitly examine the effect of ionic ordering by considering 20 to 25 randomly ordered quaternary or quinary B-site mixed compounds. \\

Ultimately, multiple types of PBE (PBE-optimized, PBEsol-optimized, PBE-D3-optimized, and PBEsol-D3-optimized) and HSE (static HSE+SOC on PBE-optimized structure) functionals are applied for subsets of all compounds being studied, and the effect of each is examined for the one or all of the following properties: the effective lattice parameter (a$_{eff}$), decomposition energy ($\Delta$H), band gap (E$_{g}$), and SLME. While the a, b, c lattice constants are computed for each cubic and non-cubic structure from every level of theory, for efficient comparison, we define the ``effective lattice parameter'' for any given pure or mixed composition material as a$_{eff}$ = (V$_{sc}$/pfu)$^{1/3}$, where V$_{sc}$ is the supercell volume and pfu is the number of perovskite formula units in the supercell. A negative $\Delta$H implies an inherent resistance of any ABX$_3$ compound to decompose to AX and BX$_2$ phases. E$_{g}$ should typically be between 1 eV and 2 eV for suitable single-junction solar absorption, whereas the SLME should be as high as possible. Equations for calculating SLME can be found in the original publications \cite{yu-2012-ident-poten} and in our past work \cite{Mannodi-HP4}. $\Delta$H is calculated using equation (1), where E$_{opt}$(S) is the total DFT energy per formula unit of any system S, k$_{B}$ is the Boltzmann constant, T is the temperature fixed to be 300K here, and x$_{i}$ is the mixing fraction of any species at A/B/X sites.

\begin{equation}\label{eq:decoE}
\begin{aligned}
    \Delta H = E_{opt}(ABX_3) - \sum_{i} x_{i}E_{opt}(AX) - \sum_{i} x_{i}E_{opt}(BX_2) + k_{B}T(\sum_{i} x_{i}ln(x_{i}))
\end{aligned}
\end{equation}

\subsection{\textit{Simulating Polymorphs using Perovskite Supercells}}

We first study 27 pure ABX$_3$ compounds, defined by A = FA, MA, or Cs, B = Pb, Sn, or Ge, and X = I, Br, or Cl, in 4 different phases, namely cubic, tetragonal, orthorhombic, and hexagonal, as pictured in \textbf{Fig. \ref{fig:structs}(a)}. All 27*4 = 108 structures are optimized using PBE, PBEsol, PBE-D3, and PBEsol-D3. We then perform HSE+SOC computations on all 108 structures using the PBE-optimized structures as input. Next, we consider 19 random alloyed compositions each in two (most likely) phases each and perform PBE as well as HSE+SOC computations on them, which leads to a total dataset of 146 points. We thus have the ability to compare the a$_{eff}$, $\Delta$H, E$_{g}$, and SLME calculated from multiple functionals for 146 compounds with any known experimental values, as well as to visualize entire datasets of different properties plotted against each other. \\

After studying the effect of DFT functional and perovskite phase, we turn our attentions to ionic ordering in mixed compounds: this is accomplished by simulating cubic MA-(Pb/Sn/Ba/Sr/Ca)-I$_3$ 4$\times$4$\times$4 supercells in quaternary (4 species mixed at B) and quinary (5 species mixed at B) compositions, with 20 possible structures each considered for the 5 quaternaries and 25 structures for the quinary. This leads to a dataset of 125 compounds, and the PBE-computed $\Delta$H and E$_{g}$ are visualized in terms of clustering of different B-site cations. Finally, we consider 6 compounds in the cubic phase, namely CsPbI$_3$, CsPbBr$_3$, CsPbCl$_3$, MAPbI$_3$, MAPbBr$_3$ and MAPbCl$_3$, and induce a series of lattice strains and octahedral distortions/rotations starting from the PBE-optimized ground state structures. First, we apply systematic compression and elongation in the lattice by changing the lattice constants and running volume-fixed geometry optimization; the changes in $\Delta$H and E$_{g}$ for the newly optimized structures are visualized against the amount of strain. Further, we distort corner-shared octahedra by changing positions of the bridging X atoms, perform geometry optimization on the new structures containing different amounts of distortion and rotation, and finally visualize the computed $\Delta$H and E$_{g}$. \\

\subsection{\textit{Compiled Datasets}}

\textbf{Table \ref{table:dataset}} presents a description of all the DFT datasets generated as part of this work, as well as the experimental data collected for comparison. Ultimately, we compute lattice constants, stability, band gap, and PV efficiency for 146 (MA-FA-Cs)(Pb-Sn-Ge)(I-Br-Cl)$_{3}$ compounds (both pure and mixed composition), from PBE, PBEsol, PBE-D3, PBEsol-D3, and HSE-PBE+SOC (henceforth referred to as HSE). 125 quinaries and quaternaries are simulated in 4$\times$4$\times$4 cubic supercells of MA-(Pb/Sn/Ba/Sr/Ca)-I$_3$ which yields lattice constants, stability, and band gaps from PBE. Lattice strain and octahedral distortion/rotation applied on (MA-Cs)(Pb)(I-Br-Cl)$_{3}$ results in datasets of 65 and 677 points respectively, with lattice constants, stability, and band gaps computed from PBE. Experimental data is collected from across several publications \cite{exp_1,exp_2,exp_3,exp_4,exp_5}, leading to lattice constants of 32 compounds, band gaps of 31 compounds, and power conversion efficiency (PCE) values of 19 compounds, which are compared against corresponding values from the multi-phase HaP datasets of 146 compounds. \\

\begin{table}[t]
    \centering
    \begin{adjustbox}{width=\textwidth}
    \begin{tabular}{|c|c|c|c|c|}
    \hline
        \textbf{Dataset} & \textbf{Chemical Space} & \textbf{Functional} & \textbf{Data Points}  &  \textbf{Properties} \\ \hline
        Multi-phase HaPs  &  (MA-FA-Cs)(Pb-Sn-Ge)(I-Br-Cl)$_{3}$  &  PBE  &  146  &  a, b, c, $\Delta$H, E$_{g}$, SLME  \\
        Multi-phase HaPs  &  (MA-FA-Cs)(Pb-Sn-Ge)(I-Br-Cl)$_{3}$  &  PBEsol  &  146  &  a, b, c, $\Delta$H, E$_{g}$, SLME  \\
        Multi-phase HaPs  &  (MA-FA-Cs)(Pb-Sn-Ge)(I-Br-Cl)$_{3}$  &  PBE-D3  &  146  &  a, b, c, $\Delta$H, E$_{g}$, SLME  \\
        Multi-phase HaPs  &  (MA-FA-Cs)(Pb-Sn-Ge)(I-Br-Cl)$_{3}$  &  PBEsol-D3  &  146  &  a, b, c, $\Delta$H, E$_{g}$, SLME  \\
        Multi-phase HaPs  &  (MA-FA-Cs)(Pb-Sn-Ge)(I-Br-Cl)$_{3}$  &  HSE-PBE+SOC  &  146  &  $\Delta$H, E$_{g}$, SLME  \\ \hline
        Quaternaries \& Quinaries  &  MA(Pb-Sn-Ba-Sr-Ca)I$_{3}$  &  PBE  &  125  &  a, b, c, $\Delta$H, E$_{g}$  \\ \hline
        Lattice strain  &  (MA-Cs)(Pb)(I-Br-Cl)$_{3}$  &  PBE  &  65  &  a, b, c, $\Delta$H, E$_{g}$  \\
        Octahedral distortion  &  (MA-Cs)(Pb)(I-Br-Cl)$_{3}$  &  PBE  &  677  &  a, b, c, $\Delta$H, E$_{g}$ \\ \hline
        Experimental Lattice Constants  &  (MA-FA-Cs)(Pb-Sn-Ge)(I-Br-Cl)$_{3}$  &  -  &  32  &  a, b, c  \\
        Experimental Band Gaps  &  (MA-FA-Cs)(Pb-Sn-Ge)(I-Br-Cl)$_{3}$  &  -  &  31  &  E$_{g}$  \\     
        Experimental PV Efficiencies  &  (MA-FA-Cs)(Pb-Sn-Ge)(I-Br-Cl)$_{3}$  &  -  &  19  &  PCE  \\  \hline       
    \end{tabular}
    \end{adjustbox}
    \caption{\label{table:dataset} Description of all the datasets used in this work, in terms of the chemical space, DFT functionals used, number of data points, and computed properties. a, b, and c refer to the optimized or experimental lattice constants. Experimental data is collected from across several publications \cite{exp_1,exp_2,exp_3,exp_4,exp_5}. PCE stands for power conversion efficiency.}
\end{table}

\begin{figure}[h]
\begin{center}
 \includegraphics[width=1.0\textwidth]{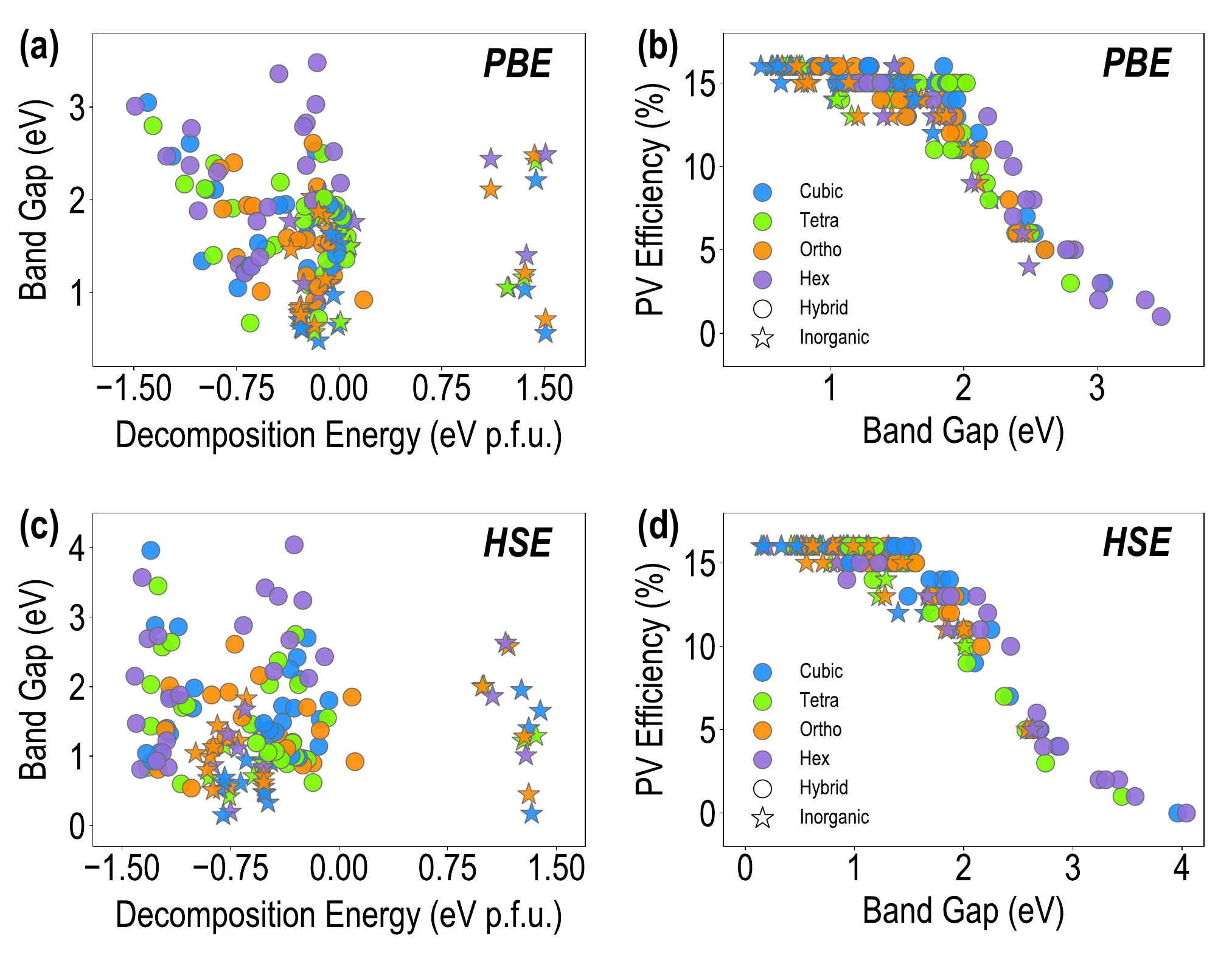}
  \caption{Visualization of the DFT dataset of multi-phase HaPs: (a) PBE $\Delta$H vs E$_{g}$, (b) PBE E$_{g}$ vs SLME (PV efficiency), (c) HSE-PBE+SOC $\Delta$H vs E$_{g}$, and (d) HSE-PBE+SOC E$_{g}$ vs SLME. Different shapes of the scatter points represent hybrid organic-inorganic HaPs and purely inorganic HaPs, and the different colors represent different perovskite phases.} \label{fig:data}
  \end{center}
\end{figure}

\section{Results and Discussion}

\subsection{\textit{Tracking Computed Properties Against Phase and Functional}}

\textbf{Fig. \ref{fig:data}} shows $\Delta$H vs E$_{g}$ and E$_{g}$ vs SLME plots for the multi-phase HaP dataset of 146 points, from PBE and HSE-PBE+SOC (henceforth referred to as HSE). Corresponding plots for PBEsol, PBE-D3, and PBEsol-D3 are presented in \textbf{Fig. S1}. The general property distributions are very similar from all 5 functionals. We note here that while the PBE-based functionals are clearly not intended to estimate band gaps and PV efficiencies, our purpose is to establish a baseline of computation accuracy from these cheaper semi-local functionals. Across the plots in \textbf{Fig. \ref{fig:data}} and \textbf{Fig. S1}, the scatter points are distinguished in terms of the perovskite phase and whether a compound is a hybrid organic-inorganic perovskite (MA- or FA-based) or a purely inorganic perovskite (Cs-based). As seen from \textbf{Fig. \ref{fig:data}(a)} and \textbf{(c)}, a majority of the compounds have $\Delta$H < 0 eV p.f.u. from both PBE and HSE, indicating a strong resistance to decomposition for nearly all chemistries and phases, except for a few inorganic compounds where $\Delta$H $\sim$ 1.5 eV p.f.u. The unstable compounds are primarily Cs-based chlorides whereas the most stable compounds (with very negative $\Delta$H) are usually FA-based iodides. There is no clear phase-specific stability preference, with all phases seemingly distributed across the range of $\Delta$H values from $\sim$ -1.5 eV p.f.u. to 1.5 eV p.f.u. \\

\begin{figure}[h]
\begin{center}
 \includegraphics[width=1.0\textwidth]{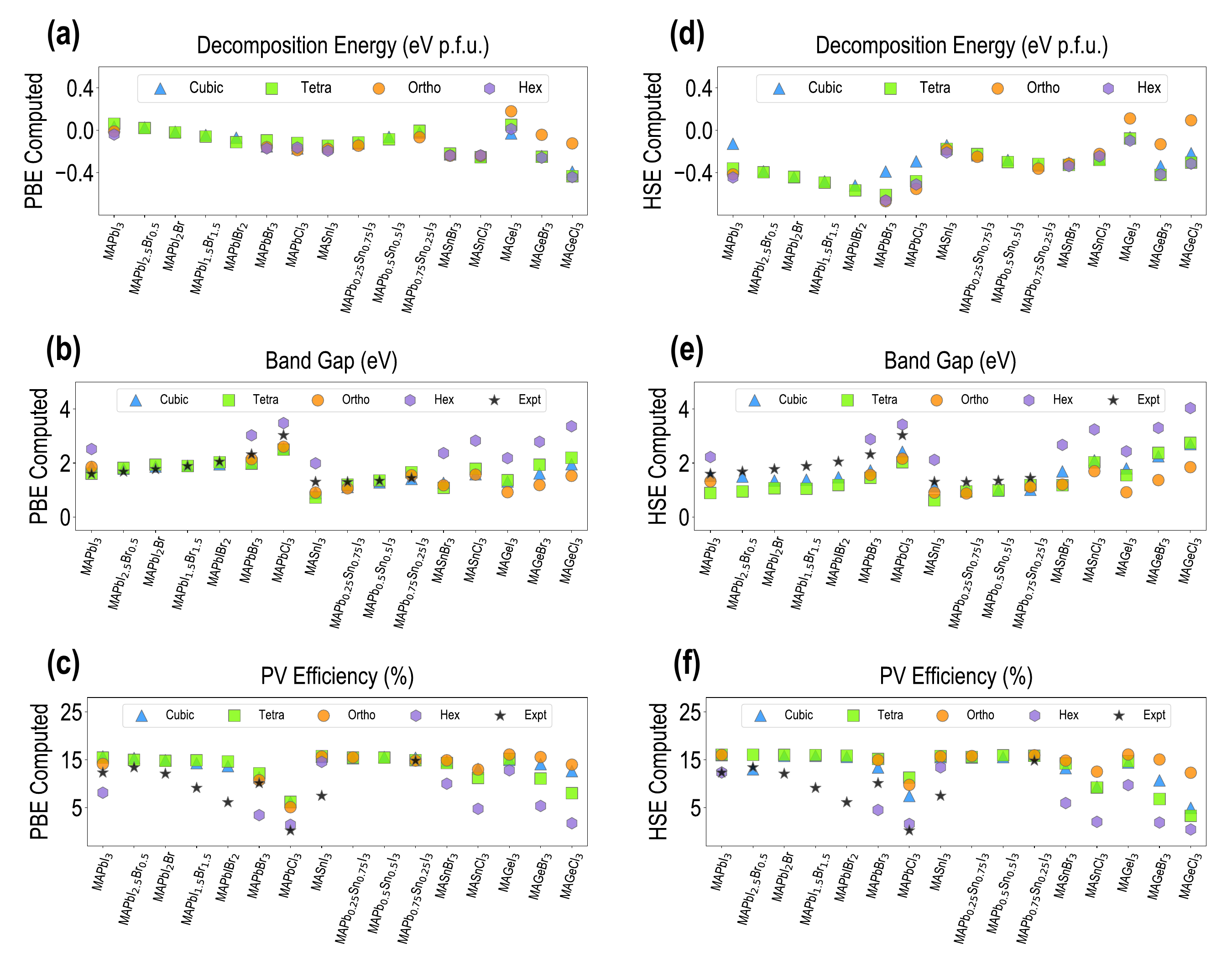}
  \caption{Multi-phase property visualization for 16 MA-based hybrid HaPs: (a) PBE $\Delta$H, (b) PBE E$_{g}$, (c) PBE SLME, (d) HSE $\Delta$H, (e) HSE E$_{g}$, and (f) HSE SLME.} \label{fig:MA_data}
  \end{center}
\end{figure}

E$_{g}$ ranges from a low of $\sim$ 0.5 eV (0.2 eV) from PBE (HSE) for cubic CsSnI$_3$ to a high of $\sim$ 3.5 eV (4 eV) from PBE (HSE) for hexagonal MAGeCl$_3$. SLME (labeled as PV efficiency in \% in the plots) values range from 0 for the highest band gap compounds to a high of 16\% from both PBE and HSE for compounds with E$_{g}$ in the vicinity of 1 eV. The SLME vs E$_{g}$ plot shows the characteristic shape that has been explored in past works \cite{slme_1,slme_2,Mannodi-HP4}. Interestingly, many of the highest SLME values are shown by cubic, tetragonal, or orthorhombic phases of Cs-based compounds, which also show some of the lowest E$_{g}$ values---with the caveat that some of these compounds also have very large $\Delta$H and are thus unstable. The outer boundary of the SLME vs E$_{g}$ plots is dominated by hexagonal phase hybrid perovskites because of their tendency to show larger E$_{g}$. Very similar ranges of values for all three properties are observed in the PBEsol, PBE-D3, and PBEsol-D3 datasets as well, as shown in \textbf{Fig. S1}, with PBEsol-D3 showing a tendency for broadening both the $\Delta$H and E$_{g}$ ranges. \textbf{Fig. S2} shows the PBE-computed $\Delta$H, E$_{g}$, and PV efficiency plotted against the corresponding values from the remaining four functionals, namely PBEsol, PBE-D3, PBEsol-D3, and HSE. We find that the $\Delta$H values track very linearly except for some compounds which are predicted to be more stable from HSE than from PBE. E$_{g}$ and PV efficiency also track linearly except for when D3 corrections are used; many of the inorganic compounds show strange behavior, seemingly arising from the fact that vdW corrections are unnecessary for Cs-based compounds although they should be incoporated for FA- and MA-based compounds. \\

\subsection{\textit{Benchmarking Computed Optoelectronic Properties}}

To further examine the phase-dependent properties of different hybrid and inorganic HaPs, we plotted the PBE and HSE properties of 16 MA-based compounds in \textbf{Fig. \ref{fig:MA_data}}, using individual plots with compound labels on the x-axis. Corresponding PBE and HSE data for 12 FA-based compounds and 17 Cs-based compounds are presented in \textbf{Figs. S3} and \textbf{S4}. Plots for the PBEsol, PBE-D3, and PBEsol-D3 data are presented across \textbf{Figs. S5}, \textbf{S6}, and \textbf{S7}. Known experimental values of E$_{g}$ and PCE are also shown in all such plots. We find that the tetragonal phase is largely the most stable one for MA compounds with the hexagonal phase also showing low energies in many cases, from both PBE and HSE. Hexagonal phase compounds show the largest E$_{g}$ and the lowest SLME, whereas the orthorhombic phase displays the opposite behavior. There is also an impressive match between the PBE-computed E$_{g}$ of tetragonal phase compounds and the corresponding measured E$_{g}$ values, which comes from the aforementioned cancelation of errors and accidental accuracy of GGA-PBE when SOC is not included \cite{Mannodi-HP4}. The HSE E$_{g}$ for the same compounds end up being under-estimated because of SOC reducing the band gap to a larger extent than desired, and likely because of a need for tuning the mixing parameter $\alpha$ in the HSE calculation. The measured PCE values on average do not match well with PBE or HSE computed SLME values, although some qualitative trends are captured. \\

\begin{figure}[htp]
\begin{center}
 \includegraphics[width=1.0\textwidth]{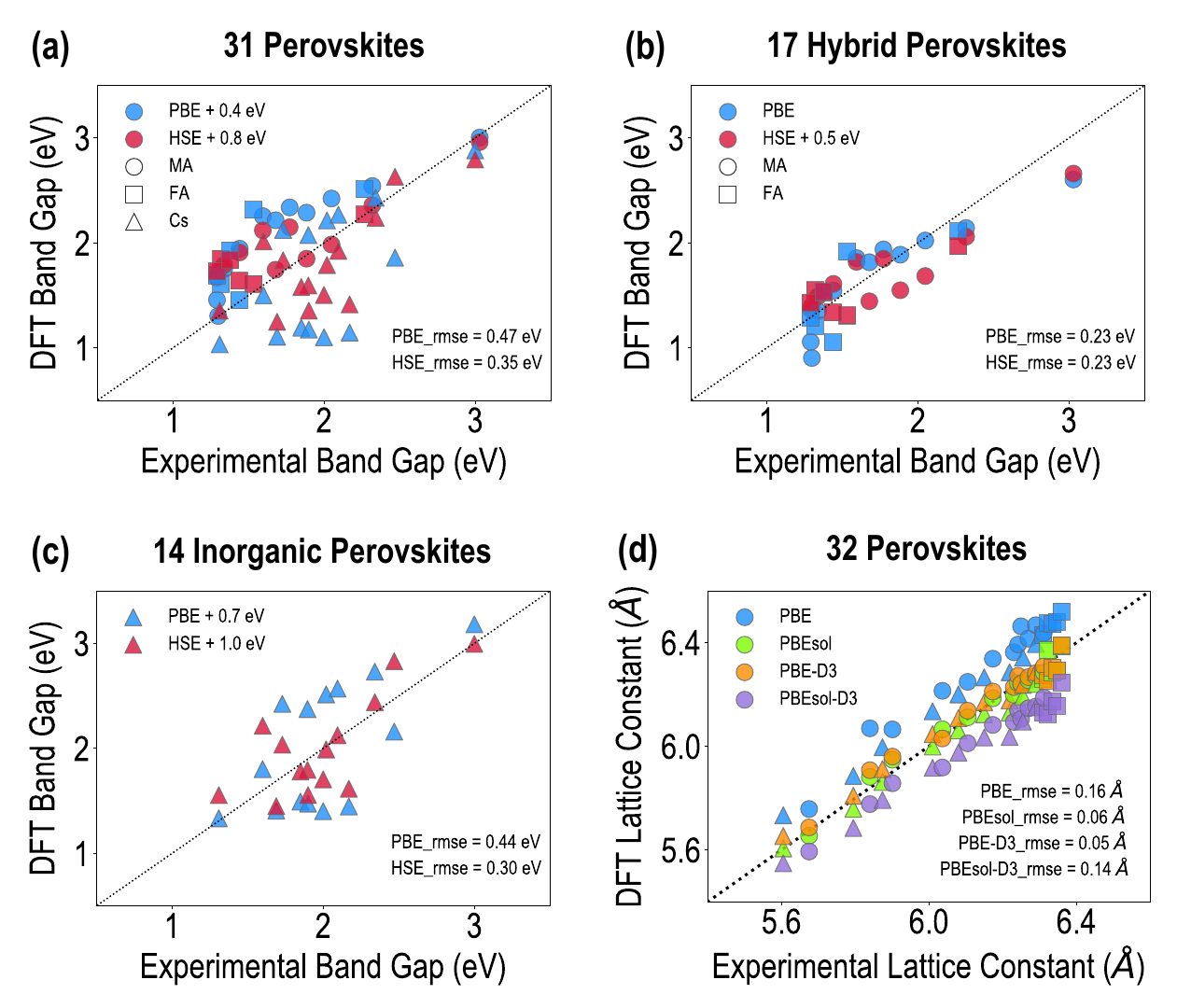}
  \caption{Accuracy of different DFT functionals compared to experiments: PBE and HSE-PBE+SOC E$_{g}$ for (a) 31 MA-, FA-, and Cs-based compounds, (b) 17 hybrid HaPs (MA- and FA-based), and (c) 14 inorganic Cs-based HaPs, and (d) pseudo-cubic lattice constants of 32 MA-, FA-, and Cs-based compounds from PBE, PBEsol, PBE-D3, and PBEsol-D3.} \label{fig:dft_expt}
  \end{center}
\end{figure}

Similarly, \textbf{Fig. S3} shows that the hexagonal or cubic phase are most preferred for FA-based compounds, there is a smaller range of E$_{g}$ values across the four phases and both PBE and HSE match very well with experiments, and the PV efficiency values from both functionals also match reasonably well with experiments. The best match with experiments for E$_{g}$ and SLME of FA compounds comes from the hexagonal phase. \textbf{Fig. S4} shows that the orthorhombic or cubic phase are most preffered for Cs compounds from both PBE and HSE, and there is little match between computed E$_{g}$ and SLME and corresponding measured values across the 17 compounds. PBE-computed E$_{g}$ and SLME show pretty good accuracy for CsPbX$_3$ compounds (where X is some combination of I, Br, and Cl) but falter for most of the Sn-containing compounds. Finally, the DFT-computed E$_{g}$ from different functionals are plotted against experimental values for 31 known compounds in \textbf{Fig. S8(a)}. Surprisingly, PBE shows the lowest root mean square error (RMSE, DFT vs experiment) of 0.57 eV, with the remaining functionals, including HSE, showing RMSE values between 0.81 eV and 1.06 eV. These errors are highly dependent on perovskite type and can certainly be improved using advanced functionals, including tuning the HSE mixing parameter. \\

To improve the DFT-experiment correspondence, we applied some corrections to the PBE and HSE E$_{g}$ values and observed a marked reduction in RMSE. \textbf{Fig. \ref{fig:dft_expt}(a)} shows PBE E$_{g}$ shifted up by 0.4 eV and HSE E$_{g}$ shifted up by 0.8 eV plotted against measured E$_{g}$ values for 31 compounds, showing RMSE values of 0.47 eV and 0.35 eV respectively for PBE and HSE. This data is plotted for 17 hybrid HaPs in \textbf{Fig. \ref{fig:dft_expt}(b)} and for 14 inorganic HaPs in \textbf{Fig. \ref{fig:dft_expt}(c)}. As observed earlier, PBE E$_{g}$ without any correction shows a very low RMSE of 0.23 eV for hybrid HaPs, while HSE E$_{g}$ shifted up by 0.5 eV also shows an RMSE of 0.23 eV. PBE E$_{g}$ shifted up by 0.7 eV improves the inorganic RMSE to 0.44 eV and the HSE E$_{g}$ shifted up by 1 eV for inorganic compounds shows an RMSE of 0.30 eV. Thus, based on the observations so far, we posit here that (a) PBE E$_{g}$ can be used with reasonable accuracy for screening across MA- and FA-based HaPs, (b) HSE+SOC E$_{g}$ (using a default mixing parameter of $\alpha$=0.25) shifted up by 1 eV can serve as an accurate estimate for Cs-based HaPs, and (c) it would very likely be possible to accurately learn the experiment-level E$_{g}$ of all important HaP compositions by combining PBE, HSE, and experimental data and performing multi-fidelity learning \cite{MF1,MF2}, as we plan to do in future work. \textbf{Table \ref{table:rmse}} lists the RMSE values for E$_{g}$ from all DFT functionals compared against experiments, with and without corrections, for different datasets. \\

\subsection{\textit{Benchmarking Computed Lattice Constants}}

Finally, we examine the accuracy and inter-dependence of lattice constant values from the four PBE-based functionals. \textbf{Fig. S9} shows the PBE-computed effective lattice parameter (a$_{eff}$) plotted against the corresponding values from PBEsol, PBE-D3, and PBEsol-D3, for the entire multi-phase HaP dataset of 146 compounds. In general, a$_{eff}$ values decrease from PBE to PBE-D3 to PBEsol to PBEsol-D3, showing how the lattice becomes more closely packed upon the inclusion of vdW interactions and the use of PBEsol. There is a clear linear correlation between a$_{eff}$ values from different functionals, and equations listed in the plots in \textbf{Fig. S9} could be trivially applied to calculate lattice parameters for any functional using the PBE values, with 99\% accuracy. It is noted here that during geometry optimization using any functional, the prototype cubic/tetragonal/orthorhombic/hexagonal structure is used as the starting point, with necessary SQS-based ionic mixing, but the cell size and shape are allowed to change a small amount for lowering the energy, meaning a lot of the compounds have slight deviations from ideal prototype structures. Lattice constants will further change if full geometry optimization were to be performed with HSE06, but as shown in our past work \cite{Mannodi-HP4}, HSE06-relaxation is often unnecessary for HaPs. \\

\begin{table}[t]
    \centering
    \begin{adjustbox}{width=0.7\textwidth}
    \begin{tabular}{|c|c|c|c|}
    \hline
        \textbf{Functional} & \textbf{Dataset} & \textbf{Property} & \textbf{RMSE}  \\ \hline
        PBE  &  32 HaPs  &  Lattice Constant ({\AA})  &  0.16  \\
        PBEsol  &  32 HaPs  &  Lattice Constant ({\AA})  &  0.06  \\
        PBE-D3  &  32 HaPs  &  Lattice Constant ({\AA})  &  0.05  \\
        PBEsol-D3  &  32 HaPs  &  Lattice Constant ({\AA})  &  0.14  \\ \hline 

        PBE  &  45 HaPs  &  Band Gap (eV)  &  0.57  \\
        PBEsol  &  45 HaPs  &  Band Gap (eV)  &  0.86  \\
        PBE-D3  &  45 HaPs  &  Band Gap (eV)  &  0.81  \\
        PBEsol-D3  &  45 HaPs  &  Band Gap (eV)  &  1.06  \\
        HSE-PBE+SOC  &  45 HaPs  &  Band Gap (eV)  &  0.83  \\ \hline

        PBE-corrected  &  45 HaPs  &  Band Gap (eV)  &  0.47  \\
        HSE-corrected  &  45 HaPs  &  Band Gap (eV)  &  0.35  \\ \hline

        PBE  &  28 Hybrid HaPs  &  Band Gap (eV)  &  0.23  \\
        HSE-corrected  &  28 Hybrid HaPs  &  Band Gap (eV)  &  0.23  \\ \hline

        PBE-corrected  &  17 Inorganic HaPs  &  Band Gap (eV)  &  0.44  \\
        HSE-corrected  &  17 Inorganic HaPs  &  Band Gap (eV)  &  0.30  \\ \hline

    \end{tabular}
    \end{adjustbox}
    \caption{\label{table:rmse} Root mean square errors between DFT-computed lattice constants and band gaps and corresponding experimental values collected from the literature.}
\end{table}

\begin{figure}[h]
\begin{center}
 \includegraphics[width=1.0\textwidth]{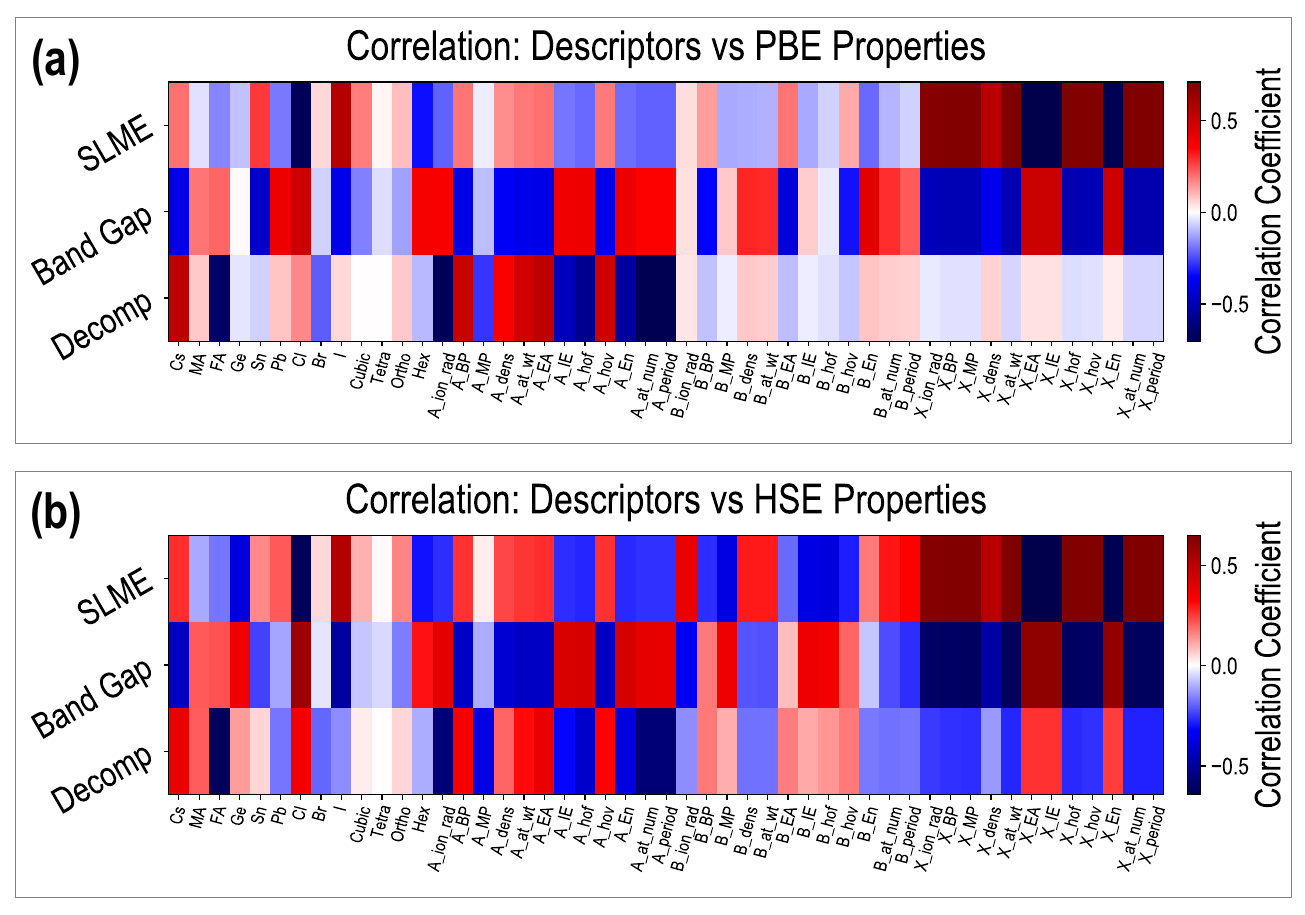}
  \caption{Pearson correlation coefficients between DFT-computed properties ($\Delta$H, E$_{g}$, and SLME) and 49-dimensional input descriptors, for (a) PBE, and (b) HSE-PBE+SOC.} \label{fig:corr}
  \end{center}
\end{figure}

\textbf{Fig. \ref{fig:dft_expt}(d)} shows the computed lattice constants for 32 selected compounds plotted against their corresponding experimentally measured values. For better visibility, the longer a/b/c lattice parameters from tetra/ortho/hex phases are moved to a separate plot in \textbf{Fig. S8(b)}. We find that all 4 functionals match reasonably well with experiments, but PBEsol and PBE-D3 show the lowest RMSE values of 0.06 {\AA} and 0.05 {\AA} respectively whereas PBE and PBEsol-D3 show higher RMSE values of 0.16 {\AA} and 0.14 {\AA} respectively. It tracks that the PBEsol corrections are desired for inorganic compounds and the D3 corrections are suitable for hybrid compounds, but both together may be unnecessary and lead to under-predicted lattice constants. PBE alone clearly over-predicts the lattice constants which motivates the inclusion of the necessary functional modifications. \textbf{Table \ref{table:rmse}} further lists the lattice constant RMSE values for all PBE-related functionals compared against experiments. \\

\subsection{\textit{Correlation Between Properties and Material Descriptors}}

Next, the DFT dataset is mined to obtain some qualitative insights into the physical and chemical factors that contribute to the HaP properties of interest, namely $\Delta$H, E$_{g}$, and SLME, computed at all levels of theory. For this purpose, we utilize a strategy applied by us in multiple previous studies, of converting every ABX$_3$ compound into unique composition-based vectorial representations \cite{Mannodi-HP3,Mannodi-HP4}. The ``descriptor'' for any compound is a 49-dimension vector, where the first 9 dimensions encode the fraction of any species (Cs, MA, FA, Ge, Sn, Pb, Cl, Br, I) in the compound using a value between 0 and 1, the next 4 dimensions provide a 1 or 0 score based on the phase of the compound, and the next 36 dimensions represent weight-averaged elemental properties of species at the A, B, and X sites, using well-known properties such as ionic radii, electronegativity, and electron affinity. \textbf{Fig. \ref{fig:corr}(a)} and \textbf{(b)} show heatmaps capturing the Pearson coefficients of linear correlation \cite{PearsonCorr} between all 49 descriptor dimensions and the three properties from PBE and HSE, respectively; corresponding plots for PBEsol, PBE-D3, and PBEsol-D3 are pictured in \textbf{Fig. S10}. \\

\begin{figure}[h]
\begin{center}
 \includegraphics[width=0.7\textwidth]{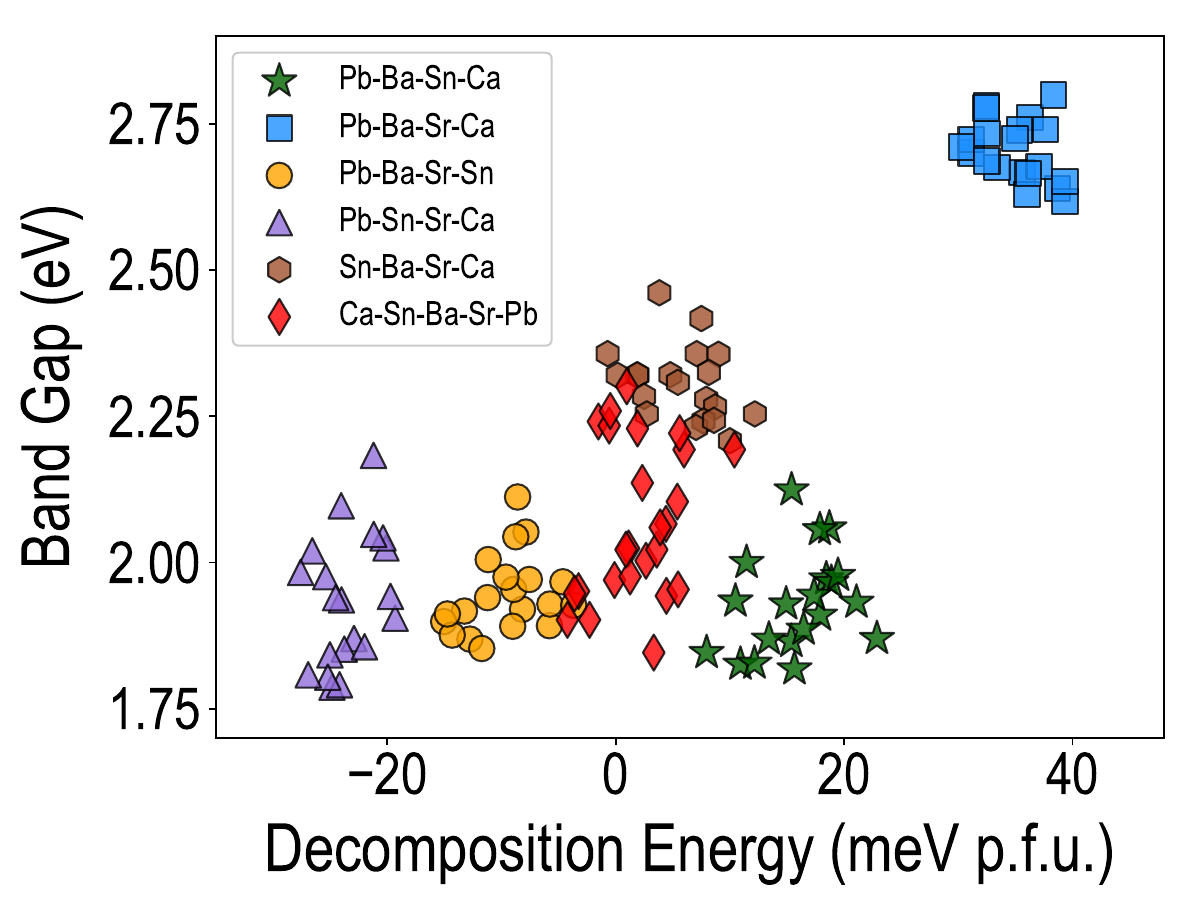}
  \caption{Effect of ionic ordering and clustering on alloy properties: PBE-computed $\Delta$H plotted against E$_{g}$ for 20 structures each of 4 quaternary compositions and 25 structures of a quinary composition.} \label{fig:large_sc_data}
  \end{center}
\end{figure}

\begin{figure}[h]
\begin{center}
 \includegraphics[width=1.0\textwidth]{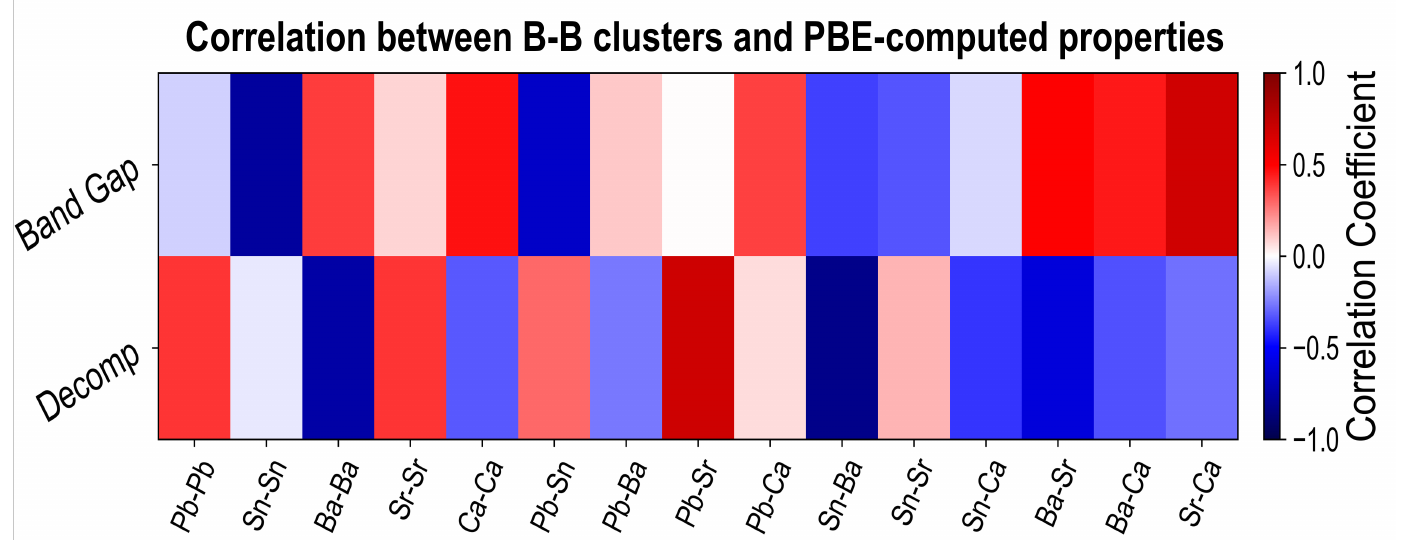}
  \caption{Pearson correlation coefficients between PBE-computed properties ($\Delta$H and E$_{g}$) and nearest neighbor pairs of B-site cations across 125 quinary and quaternary alloy structures.} \label{fig:large_sc_corr}
  \end{center}
\end{figure}

It is seen that from both PBE and HSE, the Cs-content, A-site boiling point, and A-site electron affinity have a strong positive correlation with $\Delta$H, while the FA-content, A-site ionic radius, and A-site atomic number (approximated for MA and FA using an artificial continuation of Group I, as described in past work \cite{Mannodi-HP3}) are strongly negatively correlated with $\Delta$H. This is consistent with the earlier observation that many Cs compounds have large $\Delta$H and are thus unstable, while the most stable compound with low $\Delta$H are FA-based, with MA lying somewhere in the middle. Further, the X-site properties such as ionic radius, melting point, electron affinity, and electronegativity, have the strongest correlation with both E$_{g}$ and SLME. Increasing the size, the boiling/melting point, and the heat of fusion/vaporization of the X-site species helps decrease the gap and increase the PV efficiency, while increasing the electron affinity, ionization energy, and electronegativity has the opposite effect. This is consistent the iodides or iodide-bromide compounds having the most desirable optoelectronic properties and the chlorides lying on the other end of the spectrum. The Cl-content (I-content) also shows strong negative (positive) correlation with the SLME, as shown in \textbf{Fig. \ref{fig:corr}(a)} and \textbf{(b)}. Although many B-site and A-site properties also show notable correlation with E$_{g}$ and SLME, the contributions are dominated by X-site species. Most of the qualitative correlations remain the same from the PBEsol, PBE-D3, and PBEsol-D3 datasets as well, with the inclusion of D3 corrections leading to some interesting changes in the stability trends. \\

\subsection{\textit{Effect of Ionic Ordering on HaP Alloy Properties}}

Every alloy composition investigated so far was simulated using the SQS approach in a 2$\times$2$\times$2 or 2$\times$2$\times$1 cubic or non-cubic supercell. While this is a fine representation of what the alloy would look like on average, larger supercell sizes provide the opportunity to explore different types of ordered and disordered arrangements. For this purpose, we performed additional PBE computations on a series of ``high-entropy'' HaP alloys belonging to the chemical space MA(Pb-Sn-Ba-Sr-Ca)I$_3$, as shown in \textbf{Table \ref{table:dataset}} and discussed in the Methodology section. Equimolar quinary (5 ions mixed in 20\% fractions each at the B-site) or quaternary (4 ions mixed in 25\% fractions each at the B-site) compositions would exhibit the largest mixing entropy contributions to the decomposition energy (k$_{B}$T($\sum$$_{i}$x$_{i}$ln(x$_{i}$)) in \textbf{Eqn. 1}), and are thus referred to as high-entropy perovskite alloys here. These compositions are chosen because of the general interest in MA-based iodides and in partially or completing replacing Pb in such compounds, via alloying at the B-site. We consider 4$\times$4$\times$4 cubic supercells, starting from the optimized MAPbI$_3$ geometry, and perform random mixing of ions to obtain 20 structures each for the five possible quaternaries and 25 structures for the quinary. \\

\begin{table}[!ht]
    \centering
    \begin{tabular}{|c|c|c|c|c|c|}
    \hline
        \textbf{Species} & \textbf{Pb}& \textbf{Sn}& \textbf{Ba}& \textbf{Sr}& \textbf{Ca} \\ \hline
        \textbf{Pb} & Pb-Pb & Pb-Sn & Pb-Ba & Pb-Sr & Pb-Ca\\ \hline
        \textbf{Sn} &  & Sn-Sn & Sn-Ba & Sn-Sr & Sn-Ca\\ \hline
        \textbf{Ba} &  &  & Ba-Ba & Ba-Sr & Ba-Ca\\ \hline
        \textbf{Sr} &  &  & & Sr-Sr & Sr-Ca\\ \hline
        \textbf{Ca} &  &  &  &  & Ca-Ca\\ \hline
        
    \end{tabular}
    \caption{\label{table:clusters} A matrix of possible B1-B2 pairs in all MA(Pb-Sn-Ba-Sr-Ca)I$_3$ alloys simulated in 4$\times$4$\times$4 cubic supercells.}
\end{table}

\textbf{Fig. \ref{fig:large_sc_data}} shows the PBE E$_{g}$ plotted against the PBE $\Delta$H for all 125 systems, with different colors and shapes representing different compositions. It can be seen that while the Sn-free composition MAPb$_{0.25}$Ba$_{0.25}$Sr$_{0.25}$Ca$_{0.25}$I$_{3}$ shows the largest E$_{g}$ between 2.6 eV and 2.8 eV, all other compositions show E$_{g}$ spread between $\sim$ 1.75 eV and $\sim$ 2.5 eV. The E$_{g}$ values in MAPb$_{0.25}$Sn$_{0.25}$Sr$_{0.25}$Ca$_{0.25}$I$_{3}$, for instance, range from 1.75 eV to 2.2 eV, with $\Delta$H between -30 meV p.f.u. to -18 meV p.f.u., meaning that that band gap could be changed by nearly 0.5 eV while keeping the material robustly stable against decomposition, simply by altering the ionic ordering in the system. It stands to reason that the properties displayed by the polymorphs of this material would emerge from some kind of an ensemble average over all these configurations. MAPb$_{0.25}$Sn$_{0.25}$Sr$_{0.25}$Ba$_{0.25}$I$_{3}$ also shows negative $\Delta$H across its 20 structures as E$_{g}$ ranges from $\sim$ 1.8 eV to $\sim$ 2.1 eV, implying that Pb-Sn-Sr combinations mixed with either Ba or Ca are good for achieving stability and lower gaps. \\

\begin{figure}[h]
\begin{center}
 \includegraphics[width=1.0\textwidth]{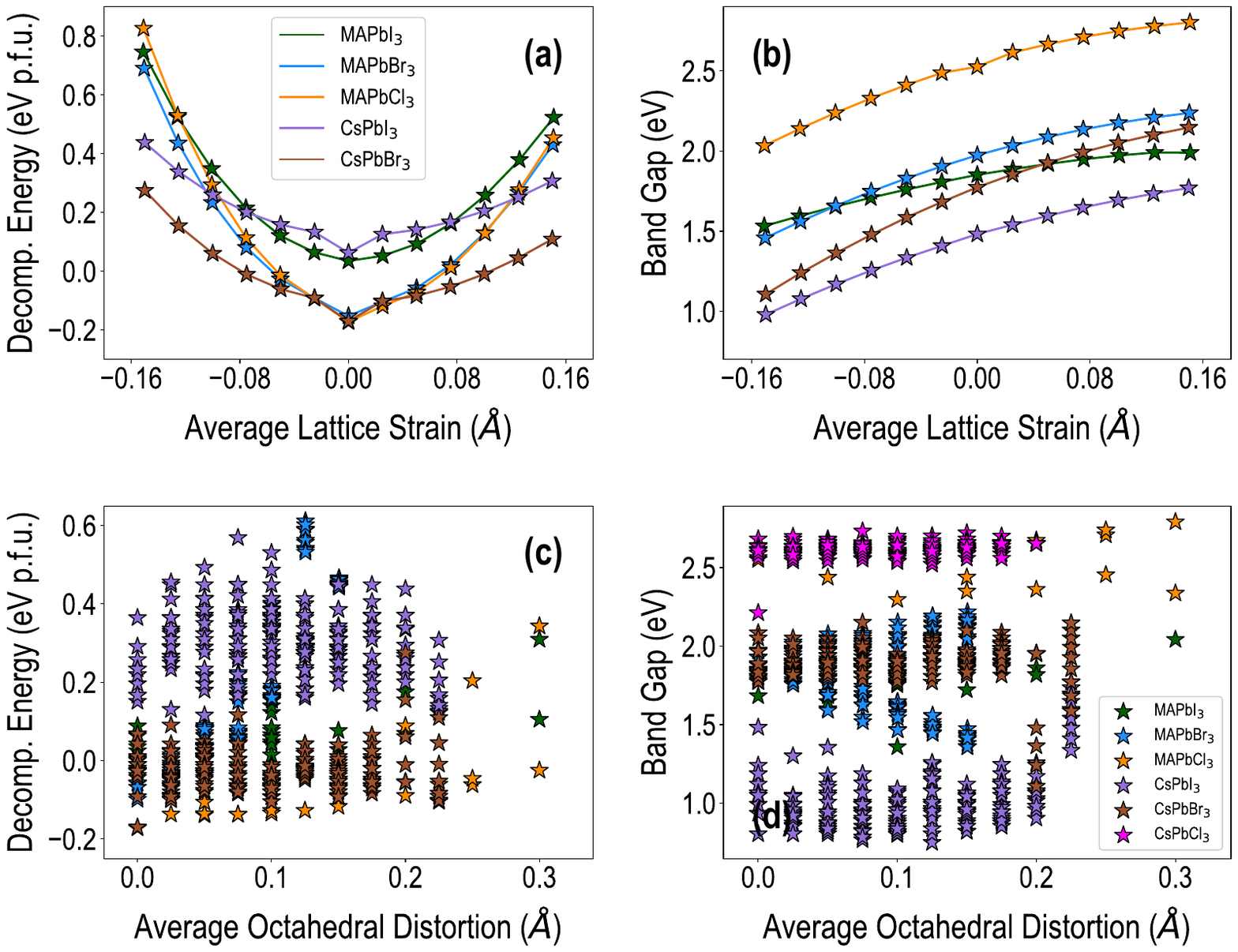}
  \caption{Effect of lattice strain and octahedral distortion/rotation for 6 selected compounds: (a) PBE $\Delta$H and (b) E$_{g}$ plotted against average lattice strain, and (c) PBE $\Delta$H and (d) E$_{g}$ plotted against average octahedral distortion. CsPbCl$_3$ is not pictured in some plots due to very large $\Delta$H values.} \label{fig:latt_oct}
  \end{center}
\end{figure}

The MAPb$_{0.2}$Sn$_{0.2}$Ba$_{0.2}$Sr$_{0.2}$Ca$_{0.2}$I$_{3}$ quinary also shows a wide range of E$_{g}$ between $\sim$ 1.8 eV and $\sim$ 2.3 eV, but $\Delta$H values become slightly positive. The same is true for the MAPb$_{0.25}$Ba$_{0.25}$Sn$_{0.25}$Ca$_{0.25}$I$_{3}$ and MASn$_{0.25}$Ba$_{0.25}$Sr$_{0.25}$Ca$_{0.25}$I$_{3}$ quaternaries, implying that Sn-Ba-Ca combinations are less desirable when it comes to the HaP stability. To further understand the effect of specific types of ionic clustering on the computed properties, we generated descriptors for all 125 structures based on the number of Pb-Pb, Pb-Sn, Sn-Sn, etc. pairs that occur in them; any B1-B2 pair is defined based on the existence of B1-I-B2 combinations with B1 and B2 connected via a bridging I anion; example structures are pictured in \textbf{Fig. \ref{fig:structs}(b)}. The matrix in \textbf{Table \ref{table:clusters}} covers all possible B1-B2 pairs, resulting in 15 possible combinations. \textbf{Fig. \ref{fig:large_sc_corr}} shows a heatmap of Pearson correlation coefficients between the 15 types of B-cation pairs and the PBE computed $\Delta$H and E$_{g}$, across the dataset of 125 quinaries and quaternaries. Interestingly, we find that Pb-Pb, Pb-Sr, and Sr-Sr pairs are least desirable for improving the perovskite stability, whereas Ba-Ba, Sn-Ba, and Ba-Sr pairs are helpful for making $\Delta$H more negative. Furthermore, Sn-Sn and Pb-Sn pairs are most responsible for reducing E$_{g}$, which is consistent with most Sn-based compounds showing lower gaps than Sn-free compounds, and pairs such as Ba-Ba, Ca-Ca, Ba-Sr, and Sr-Ca help increase the E$_{g}$. Overall, this analysis reveals that (a) by changing the ionic ordering in a given HaP composition, E$_{g}$ could be drastically reduced or increased while keeping the material stable, and (b) certain B-site cations clustering together may be helpful or harmful to the stability and desired E$_{g}$. \\

\subsection{\textit{Effect of Lattice Strain and Octahedral Distortion and Rotation on HaP Properties}}

Finally, we investigate how intentional distortions applied upon the perovskite lattice and/or on the corner-shared BX$_6$ octahedra can change the stability and band gap. For this analysis, 6 compounds given by the chemical space (MA-Cs)(Pb)(I-Br-Cl)$_3$ are considered, as listed in \textbf{Table \ref{table:dataset}} and described in the Methodology section. \textbf{Fig. \ref{fig:latt_oct}(a)} and \textbf{(b)} show the PBE $\Delta$H and PBE E$_{g}$ respectively plotted against the average lattice strain in {\AA} (over x, y, and z dimensions), for 5 compounds. CsPbCl$_3$ structures are missing in these plots because of their much higher $\Delta$H. It can be seen from \textbf{Fig. \ref{fig:latt_oct}(a)} that $\Delta$H shows an expected trend, becoming more positive for both negative (compression) and positive (elongation) lattice strain, with the lowest energy shown by the no-strain structures as expected. Interestingly, E$_{g}$ generally shows a monotonically increasing behavior all the way from large negative to large positive lattice strain. Once again, it should be noted that compounds such as CsPbBr$_3$ and MAPbBr$_3$ can be kept metastable with small amounts of strain, as shown by the negative $\Delta$H values, while changing the E$_{g}$ by nearly 0.3 eV. \\

\begin{figure}[h]
\begin{center}
 \includegraphics[width=0.7\textwidth]{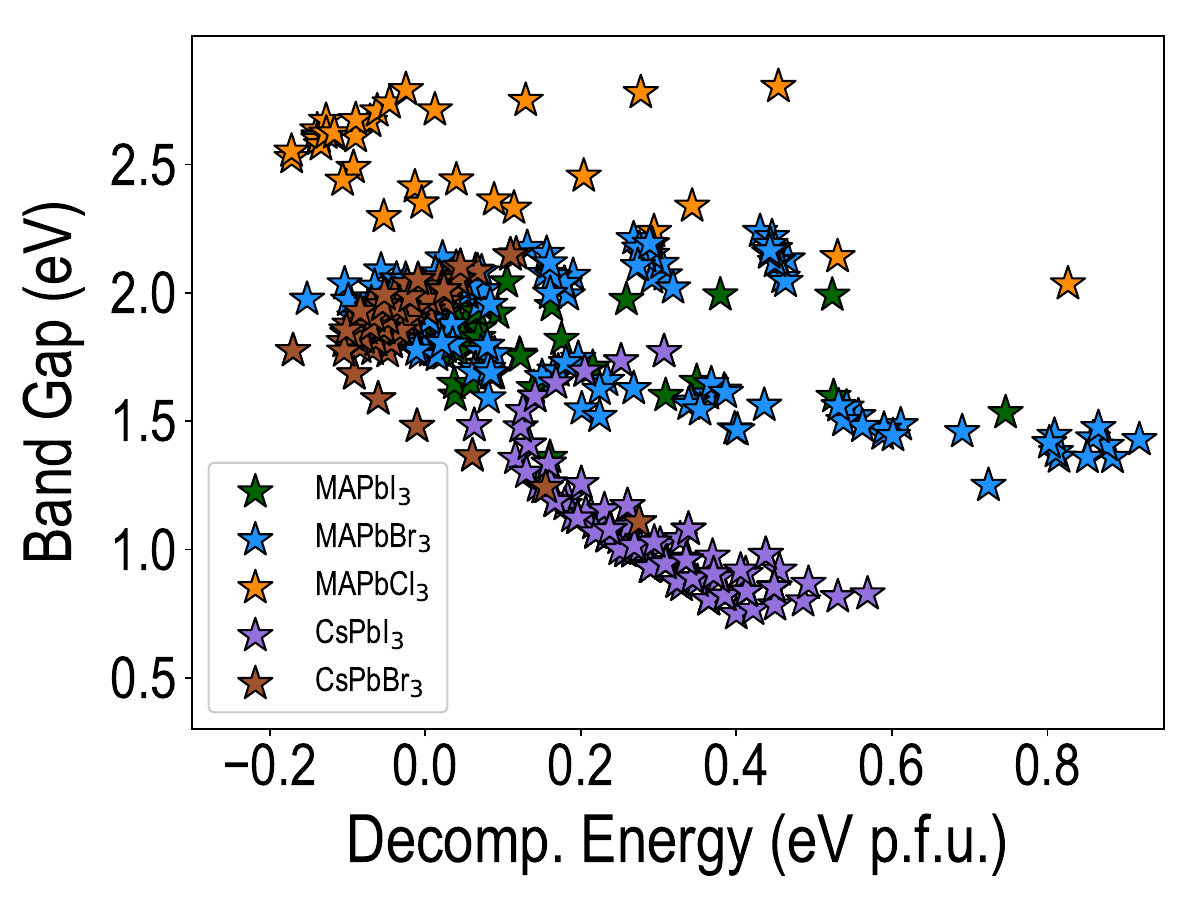}
  \caption{PBE-computed E$_{g}$ plotted against $\Delta$H for the dataset of strained and octahedrally manioulated HaPs.} \label{fig:oct_data}
  \end{center}
\end{figure}

\textbf{Fig. \ref{fig:latt_oct}(c)} and \textbf{(d)} show the PBE $\Delta$H and PBE E$_{g}$ respectively plotted against the average octahedral distortion in {\AA}. It can be seen that there are many configurations of CsPbBr$_3$, MAPbBr$_3$, and MAPbCl$_3$ with small amounts of octahedral distortion that maintain $\Delta$H below 0 eV p.f.u. and change E$_{g}$ by as much as 0.5 eV in some cases. Distorted CsPbCl$_3$ has a high E$_{g}$ $>$ 2.5 eV which does not change by a great amount as the average distortion increases from $\sim$ 0.15 {\AA} to 0.3 {\AA}, and also shows large positive $\Delta$H. Comparing plots in \textbf{Fig. \ref{fig:latt_oct}(c)} and \textbf{(d)} to plots in \textbf{Fig. \ref{fig:latt_oct}(a)} and \textbf{(b)}, it is clear that is a non-trivial correlation between octahedral distortion and the properties as compared to lattice strain vs the properties. Ideally, one would express the octahedral distortion in terms of how every atom in a BX$_6$ octahedral unit is displaced relative to all other BX$_6$ units it is connected to, and correlations would be sought between an ``octahedral distortion vector'' and the properties of interest. With the current representation, a range of property values are often observed for the same average distortion, which arises from different types of distortions resulting in lowering or raising $\Delta$H or E$_{gap}$. To wrap up this discussion, we plotted the entire dataset of lattice/octahedra strained/distorted structures for all compounds in \textbf{Fig. \ref{fig:oct_data}}, in terms of the PBE E$_{g}$ against the PBE $\Delta$H. By drawing a cut-off at a low enough $\Delta$H value, dozens of possible structures could be obtained for the same composition, showing a range of E$_{g}$ values that may be suitable for different applications. \\

\section{Perspective and Future Work}

Our systematic first principles investigation reveals that the lattice parameters, energetic stability, and optoelectronic properties of ABX$_3$ halide perovskites depend heavily on the identity and mixing of atoms at the A, B, or X sites, the perovskite phase, the level of theory being used, the type of ionic ordering in an alloy, and on possible lattice strains and octahedral distortion or rotation. Ideally, a framework for predicting the properties of HaPs, and designing novel HaP compositions/structures with multiple targeted properties, must take each of these factors into account, potentially in addition to important experimental conditions and past experimental measurements. Generating large DFT datasets as in the present work is invaluable for meaningful correlations and inter-dependencies between different properties and different levels of theory, as well as for establishing a reliable benchmark for computations against experiments. An important question that will be raised is in the usefulness of different PBE and HSE functionals going forward, when applying them to other related HaP compositions and structures. Other than applying simple linear corrections like pictured in \textbf{Fig. \ref{fig:dft_expt}}, the best way to improve DFT-level predictions is to continue applying higher levels of theory and tuning necessary parameters in a compound-by-compound case, which is clearly not conducive to a high-throughput treatment. \\

\begin{figure}[h]
\begin{center}
 \includegraphics[width=0.7\textwidth]{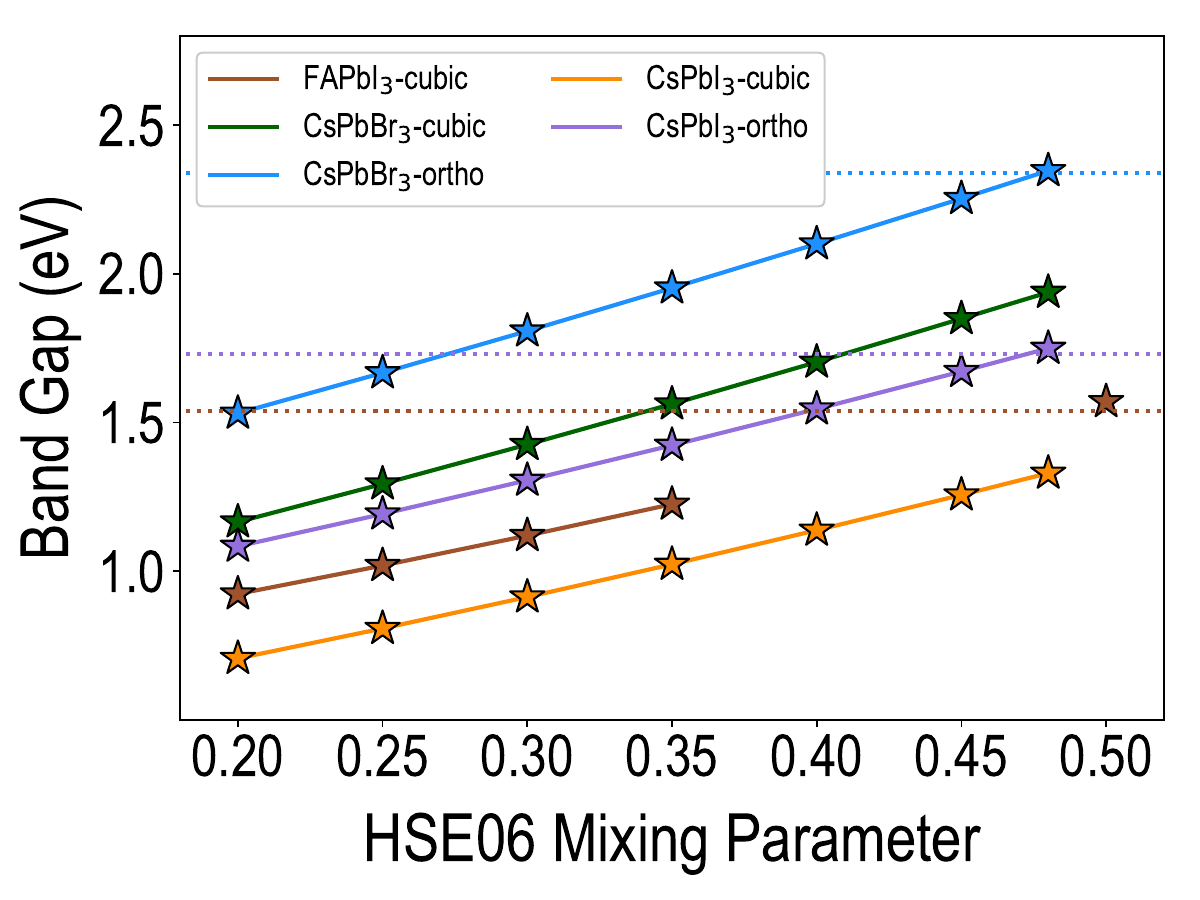}
  \caption{Band gaps computed from HSE-PBE+SOC for 5 compounds, plotted as a function of the HSE06 mixing parameter ($\alpha$). The dotted horizontal lines represent known experimental values.} \label{fig:HSE_alpha}
  \end{center}
\end{figure}

As an example, we performed new HSE+SOC computations for 5 compounds by tuning the mixing parameter $\alpha$ all the way from 0.20 to 0.50; these results are shown in \textbf{Fig. \ref{fig:HSE_alpha}}, along with known experimental values. We find that $\alpha$=0.50 reproduces the experimental E$_{g}$ perfectly for cubic FAPbI$_3$, and $\alpha$=0.48 work best for orthorhombic CsPbI$_3$ and CsPbBr$_3$. The same values of $\alpha$ might work for cubic CsPbI$_3$ and CsPbBr$_3$ as well. Such an analysis may be enormously useful for these particular compounds, as HSE+SOC computations could now be performed using ideal $\alpha$ values for optical absorption or defect calculations. Of course, it must be noted that the input structure used for these static HSE+SOC computations is itself a huge factor; for \textbf{Fig. \ref{fig:HSE_alpha}}, we utilized the FAPbI$_3$ and CsPb(I/Br)$_3$ structures that matched best with experiments, from across the different PBE-based functionals discussed earlier. Once again, it should be noted that HSE+SOC with default $\alpha$ provides means for uniform evaluation of properties across a large chemical space, whereas tuning would need to be performed independently for every compound. Machine learning (ML) models could potentially be trained in the future that combine data from different PBE and HSE functionals to yield the ideal $\alpha$ values for any given HaP composition/structure. \\

Currently, all the DFT data presented in this work is being utilized for training a myriad of ML predictive models, using both composition-based descriptors \cite{Mannodi-HP3} explained earlier in the article and used extensively in the past, as well as using state-of-the-art crystal graph-based neural network (GNN) models. While the former are elegant and simple models, they cannot typically be applied for multiple polymorphs of the same composition, whereas GNN models can appropriately represent entire crystal structures as graphs and use them as input to train predictive NN models that yield multiple properties as output \cite{CGCNN,MEGNET,ALIGNN}. A crystal graph representation automatically takes into account the identity, mixing, and bond-lengths between atoms, any lattice or octahedral distortion, ionic ordering, perovskite phase, etc. Furthermore, in addition to material $\rightarrow$ property forward predictive models, inverse design models could also be trained using techniques such as genetic algorithm \cite{GA}, Bayesian optimization \cite{BO}, generative neural networks \cite{GAN}, and variational autoencoders \cite{VAE-1}, to design novel HaP atom-composition-structure combinations that show the desired mix of negative $\Delta$H, PV-suitable E$_{g}$, and highest possible PV efficiency. \\

\section{Conclusion}

In conclusion, we performed a series of first principles-based DFT computations to investigate polymorphism in halide perovskites, in terms of changing the perovskite composition and phase, ionic ordering in alloys, and strain/distortions applied to the lattice and interconnected octehdral units. Our work shows that the stability, band gap, and PV efficiency can vary quite a lot for the same composition when the phase or ionic ordering is changed, or when distortions are applied. Different semi-local and non-local DFT functionals are explored, revealing that PBEsol or PBE-D3 may be necessary for estimating correct lattice parameters, whereas PBE and/or HSE+SOC computed band gaps can match very well with experiments if some corrections are applied. We further find that different types or ordering or distortion could be used to keep materials stable while drastically altering their band gaps. Our linear correlation analyses further show the positive or negative effect of different cations/anions, their elemental properties, and their clustering, on the bulk properties of interest. All computational data is made available to the community and is currently being utilized for trained multiple machine learning models and for guiding collaborative experimental synthesis and characterization. \\

\section*{Conflicts of Interest}
There are no conflicts to declare.

\section*{Data Availability}
All tabulated data and scripts used to analyze the DFT computed properties can be accessed from https://github.com/mannodiarun/perovs$\_$mfml$\_$ga/tree/polymorph$\_$data. \\
Crystal structure files for all datasets discussed in this article are attached with the Supplementary Information. \\

\section*{Acknowledgements}
This work was performed at Purdue University, under startup account F.10023800.05.002 from the Materials Engineering department. This research used resources of the National Energy Research Scientific Computing Center (NERSC), the Laboratory Computing Resource Center (LCRC) at Argonne National Laboratory, and the Rosen Center for Advanced Computing (RCAC) clusters at Purdue. \\

\bibliography{mybibfile}

\clearpage
\newpage
\pagenumbering{gobble}
\thispagestyle{empty} 

\onecolumn

\setcounter{figure}{0}   
\setcounter{table}{0} 
\renewcommand{\thetable}{S\Roman{table}} 
\renewcommand\thefigure{S\arabic{figure}}

\begin{center}
\vspace{0.5cm}
\Large
\textbf{Supplemental material to "First Principles Investigation of Polymorphism in Halide Perovskites"\\}
\vspace{0.5cm}
\large

\noindent\large{Jiaqi Yang\textsuperscript{a} and Arun Mannodi-Kanakkithodi\textsuperscript{a}}\\

\vspace{0.3cm}

\normalsize

  \textsuperscript{a}School of Materials Engineering, Purdue University, West Lafayette, IN 47907, USA; E-mail: amannodi@purdue.edu

\end{center}

\footnote{
\textsuperscript{a}amannodi@purdue.edu\hspace{0.3cm}}

\begin{figure}[h]
\begin{center}
 \includegraphics[width=1.0\textwidth]{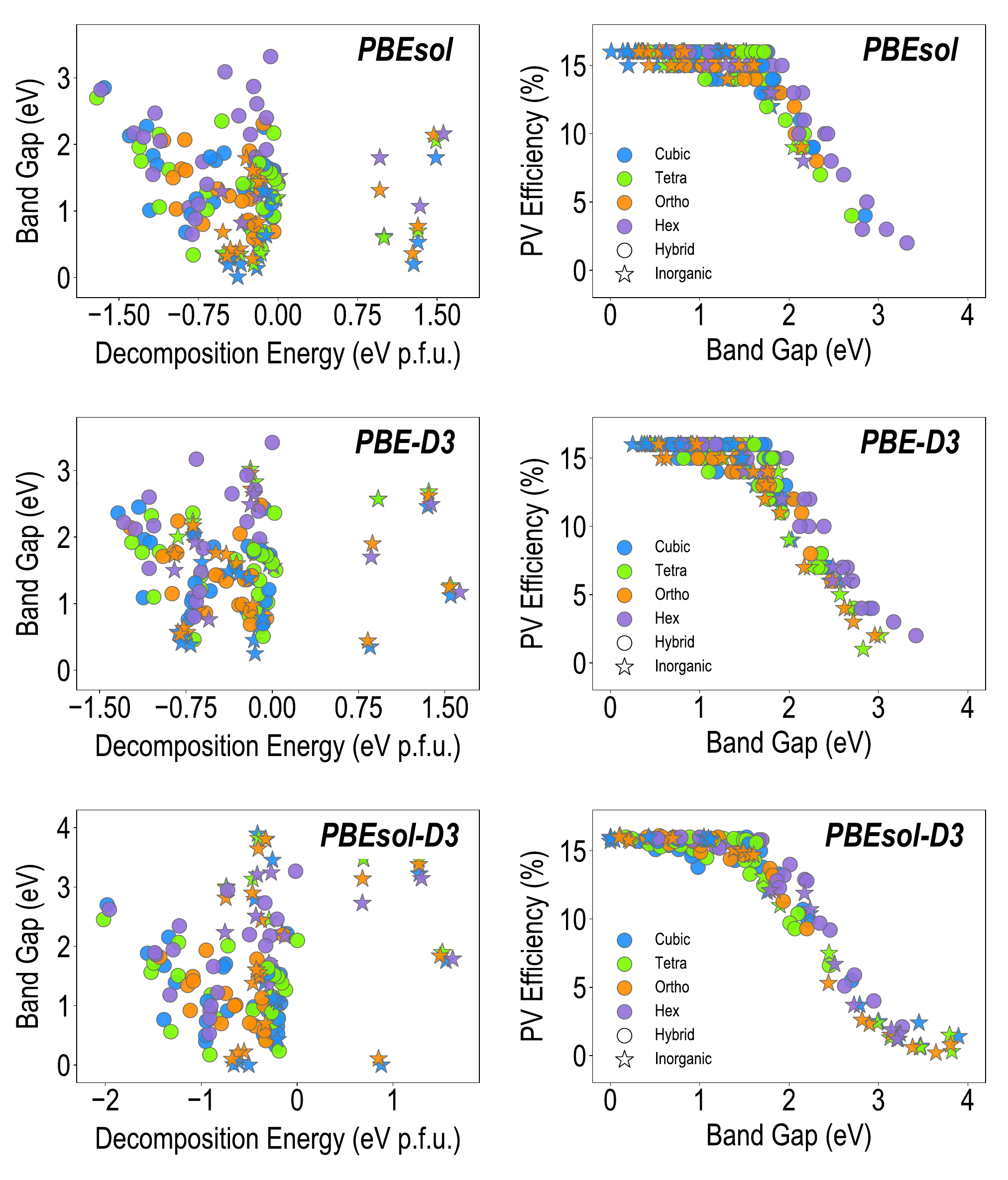}
  \caption{Visualization of the PBEsol, PBE-D3, and PBEsol-D3 datasets in terms of $\Delta$H vs E$_{g}$ and E$_{g}$ vs SLME plots.} \label{fig:S1}
  \end{center}
\end{figure}

\begin{figure}[h]
\begin{center}
 \includegraphics[width=1.0\textwidth]{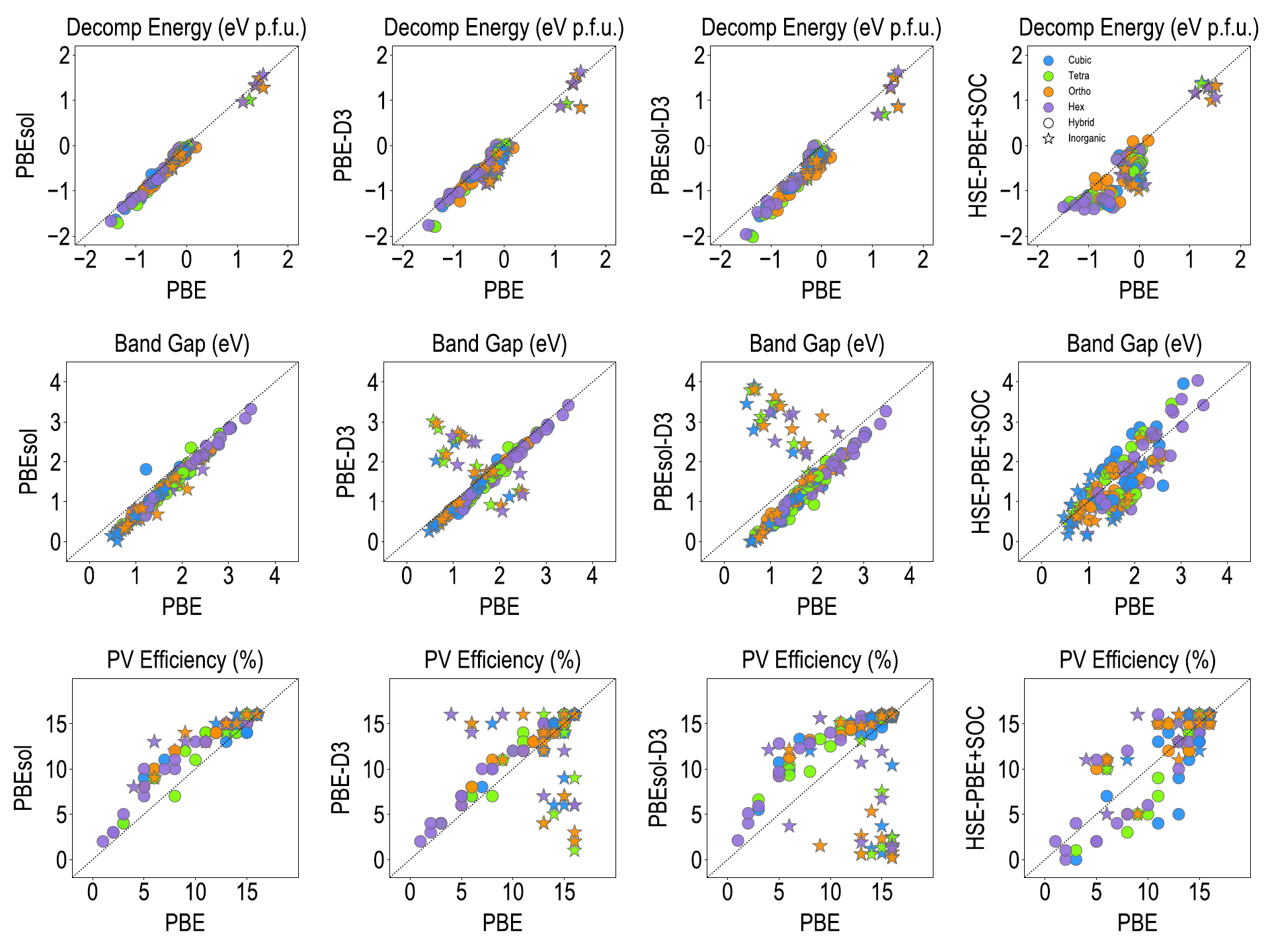}
  \caption{PBE-computed $\Delta$H, E$_{g}$, and SLME plotted against corresponding values from PBEsol, PBE-D3, PBEsol-D3, and HSE-PBE+SOC.} \label{fig:S2}
  \end{center}
\end{figure}

\begin{figure}[h]
\begin{center}
 \includegraphics[width=1.0\textwidth]{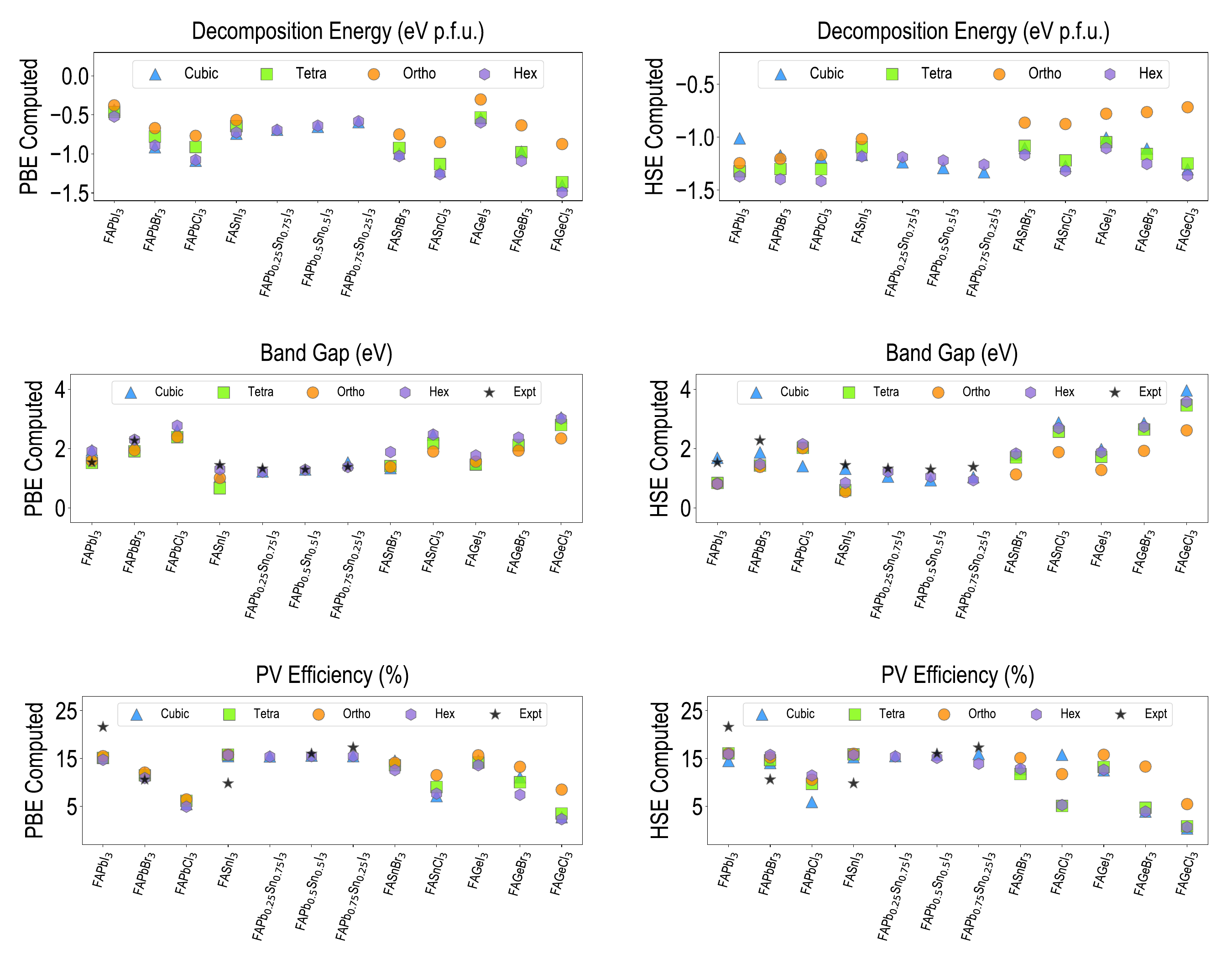}
  \caption{Multi-phase property visualization for 12 FA-based hybrid HaPs, showing $\Delta$H, E$_{g}$, and SLME from PBE and HSE-PBE+SOC.} \label{fig:S3}
  \end{center}
\end{figure}

\begin{figure}[h]
\begin{center}
 \includegraphics[width=1.0\textwidth]{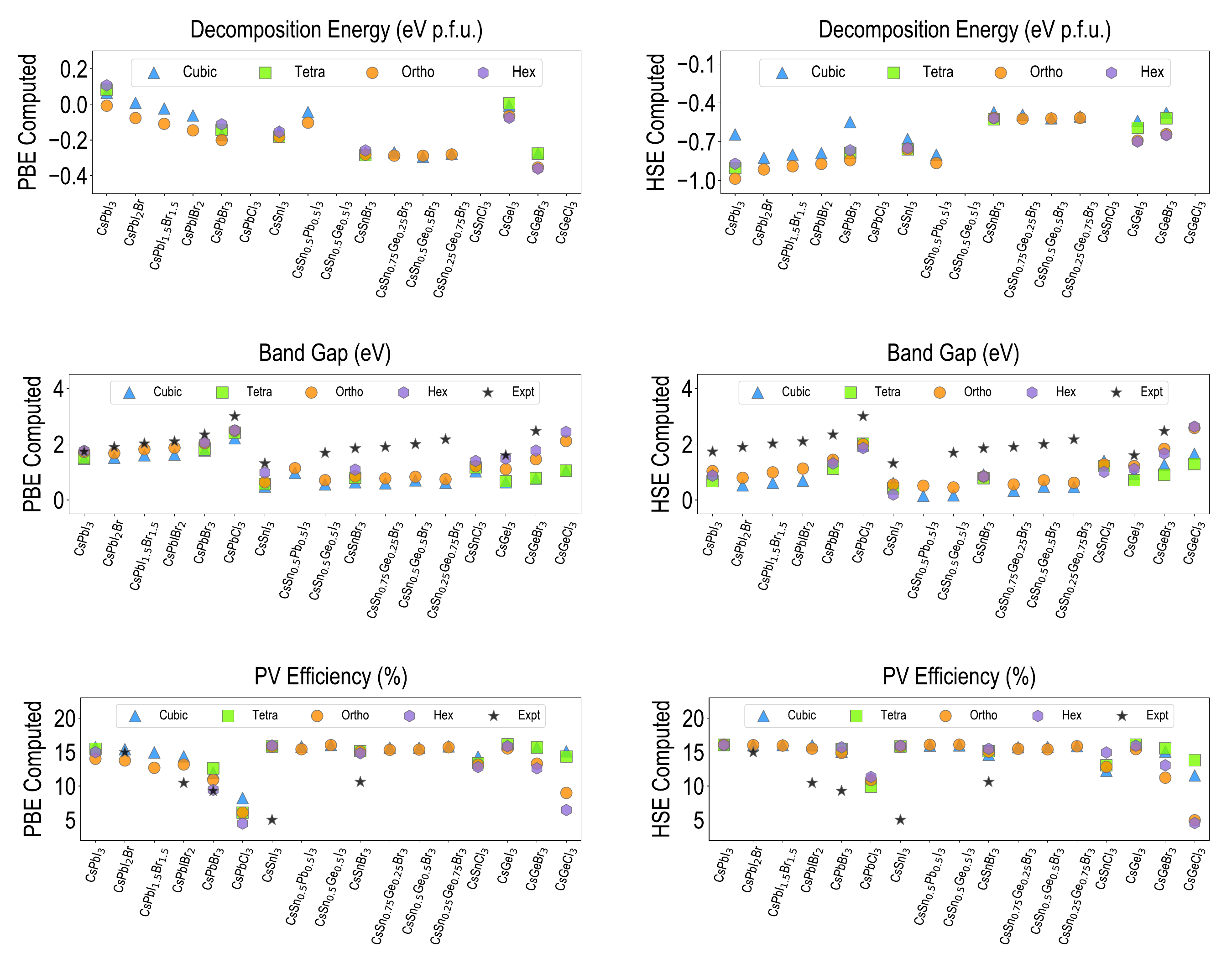}
  \caption{Multi-phase property visualization for 27 Cs-based inorganic HaPs, showing $\Delta$H, E$_{g}$, and SLME from PBE and HSE-PBE+SOC.} \label{fig:S4}
  \end{center}
\end{figure}

\begin{figure}[h]
\begin{center}
 \includegraphics[width=1.0\textwidth]{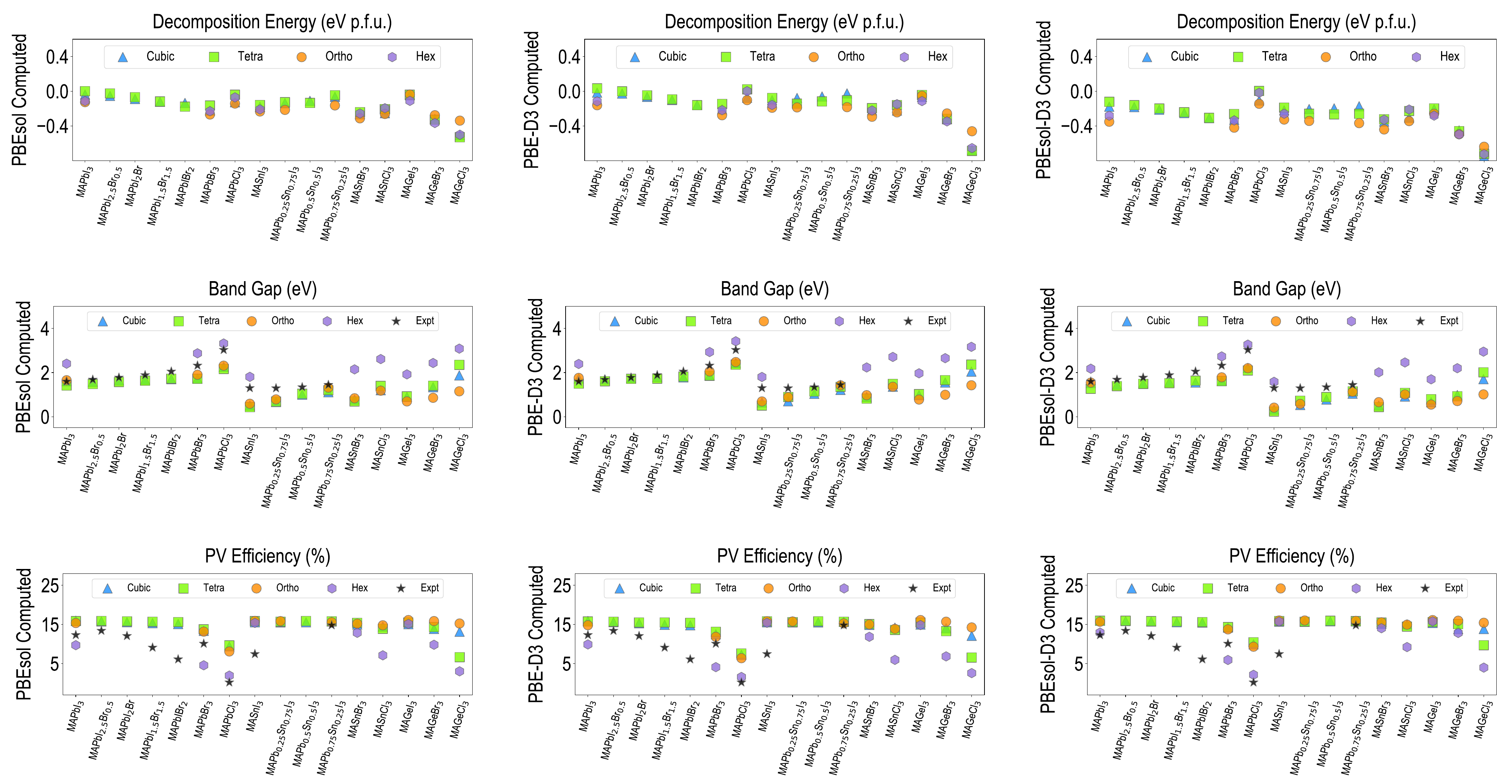}
  \caption{Multi-phase property visualization for 16 MA-based hybrid HaPs, showing $\Delta$H, E$_{g}$, and SLME from PBEsol, PBE-D3, and PBEsol-D3.} \label{fig:S5}
  \end{center}
\end{figure}

\begin{figure}[h]
\begin{center}
 \includegraphics[width=1.0\textwidth]{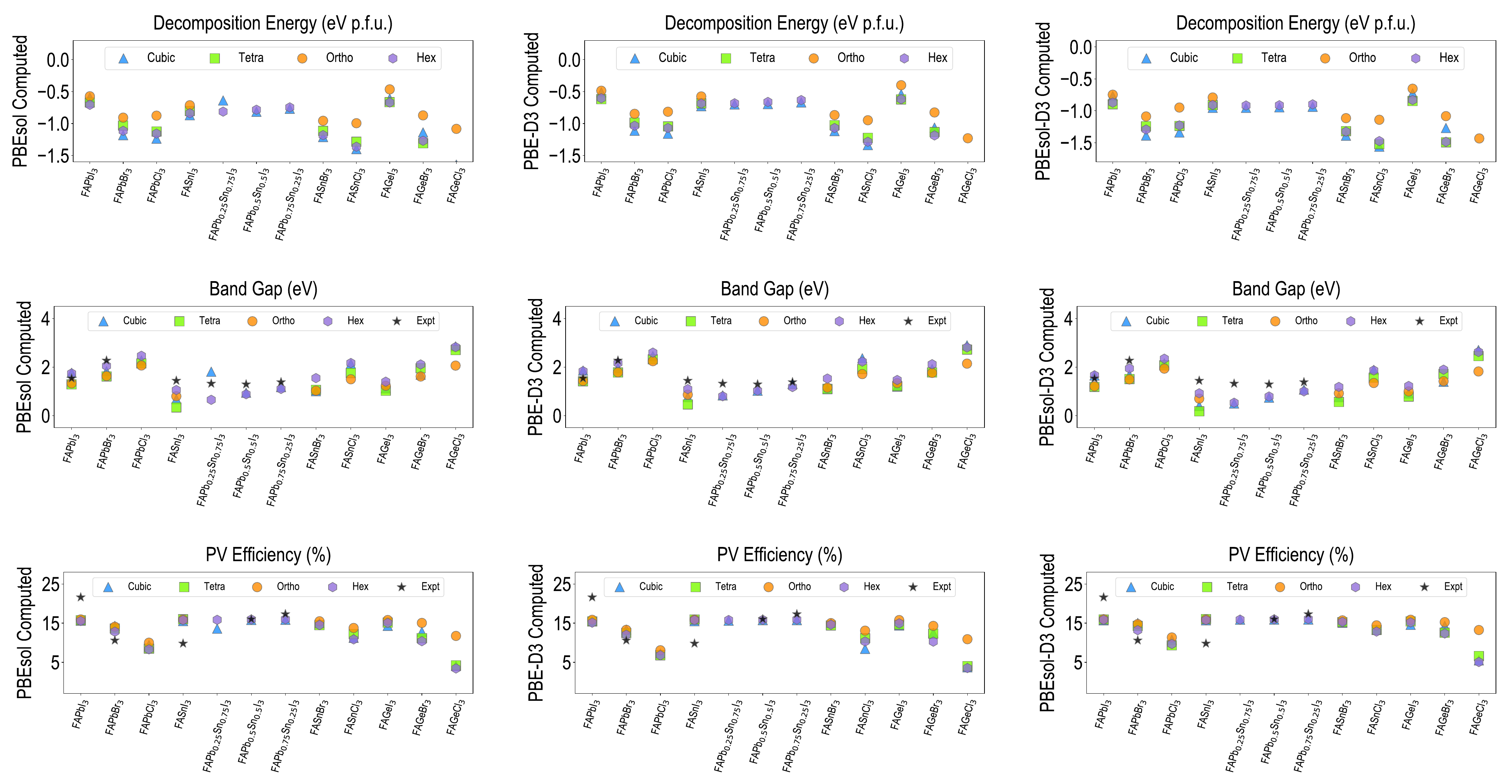}
  \caption{Multi-phase property visualization for 12 FA-based hybrid HaPs, showing $\Delta$H, E$_{g}$, and SLME from PBEsol, PBE-D3, and PBEsol-D3.} \label{fig:S6}
  \end{center}
\end{figure}

\begin{figure}[h]
\begin{center}
 \includegraphics[width=1.0\textwidth]{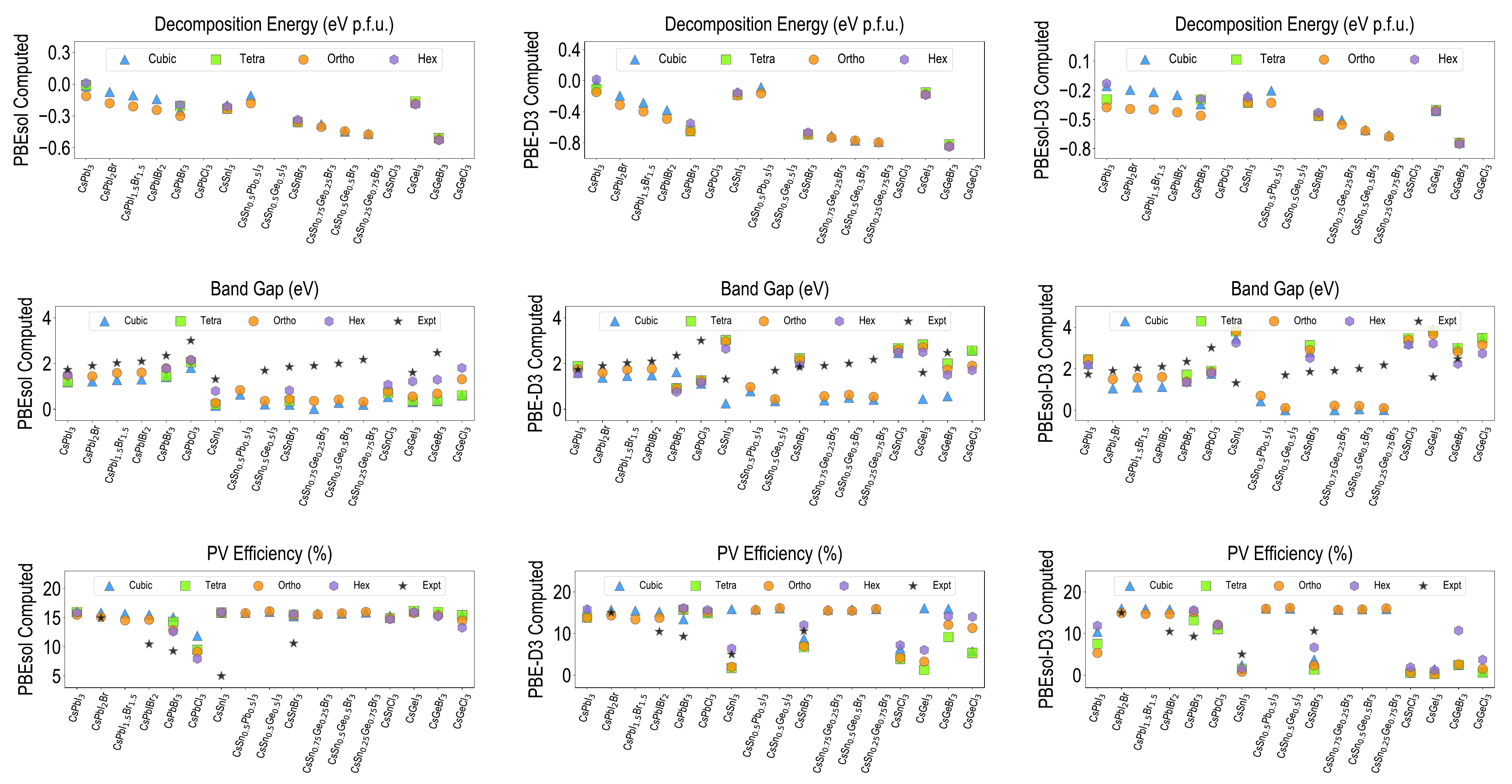}
  \caption{Multi-phase property visualization for 27 Cs-based inorganic HaPs, showing $\Delta$H, E$_{g}$, and SLME from PBEsol, PBE-D3, and PBEsol-D3.} \label{fig:S7}
  \end{center}
\end{figure}

\begin{figure}[h]
\begin{center}
 \includegraphics[width=1.0\textwidth]{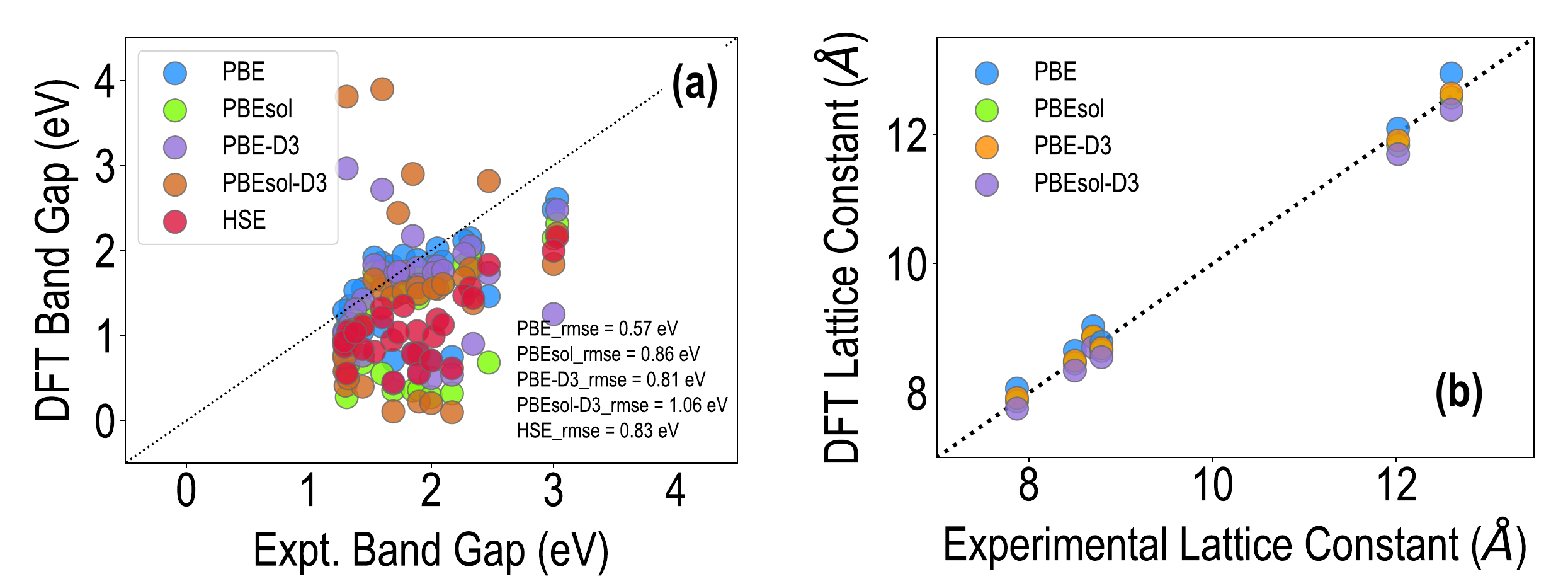}
  \caption{(a) DFT-computed band gaps plotted against measured values for 45 compounds, showing values from PBE, PBEsol, PBE-D3, PBEsol-D3, and HSE-PBE+SOC. (b) DFT-computed non-cubic lattice constants plotted against measured values for 6 compounds, showing values from PBE, PBEsol, PBE-D3, and PBEsol-D3.} \label{fig:S8}
  \end{center}
\end{figure}

\begin{figure}[h]
\begin{center}
 \includegraphics[width=1.0\textwidth]{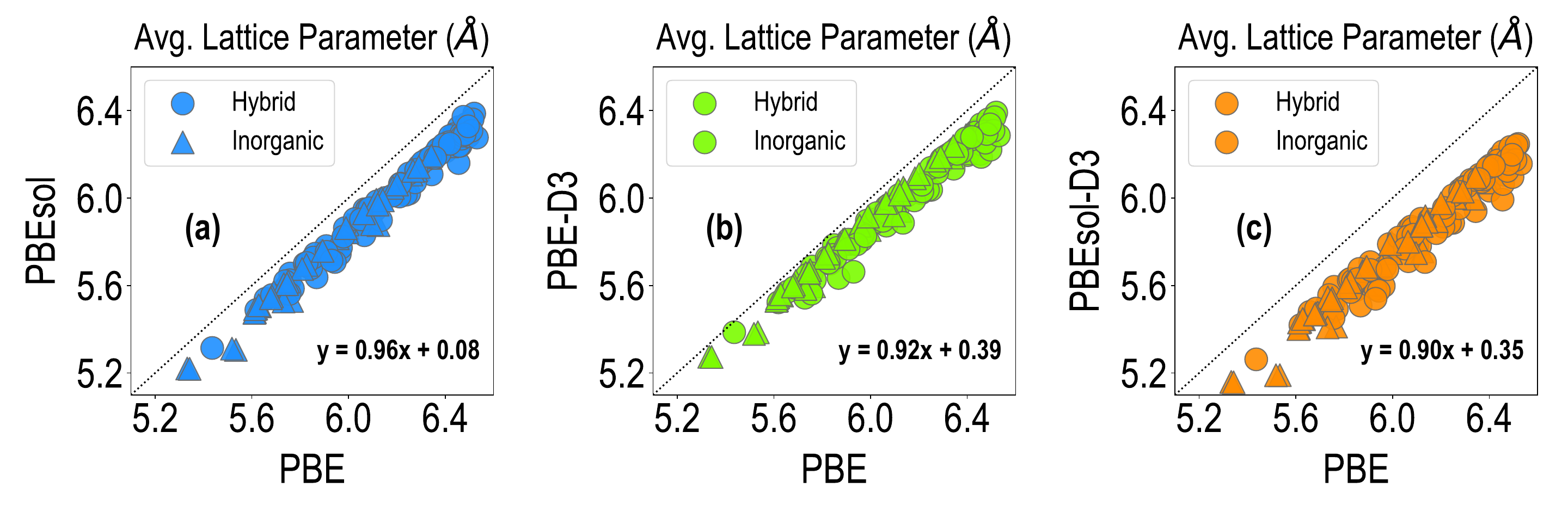}
  \caption{Average lattice parameter, defined as the cube root of the supercell volume per formula unit in any perovskite structure, plotted from PBE against corresponding values from PBEsol, PBE-D3, and PBEsol-D3.} \label{fig:S9}
  \end{center}
\end{figure}

\begin{figure}[h]
\begin{center}
 \includegraphics[width=1.0\textwidth]{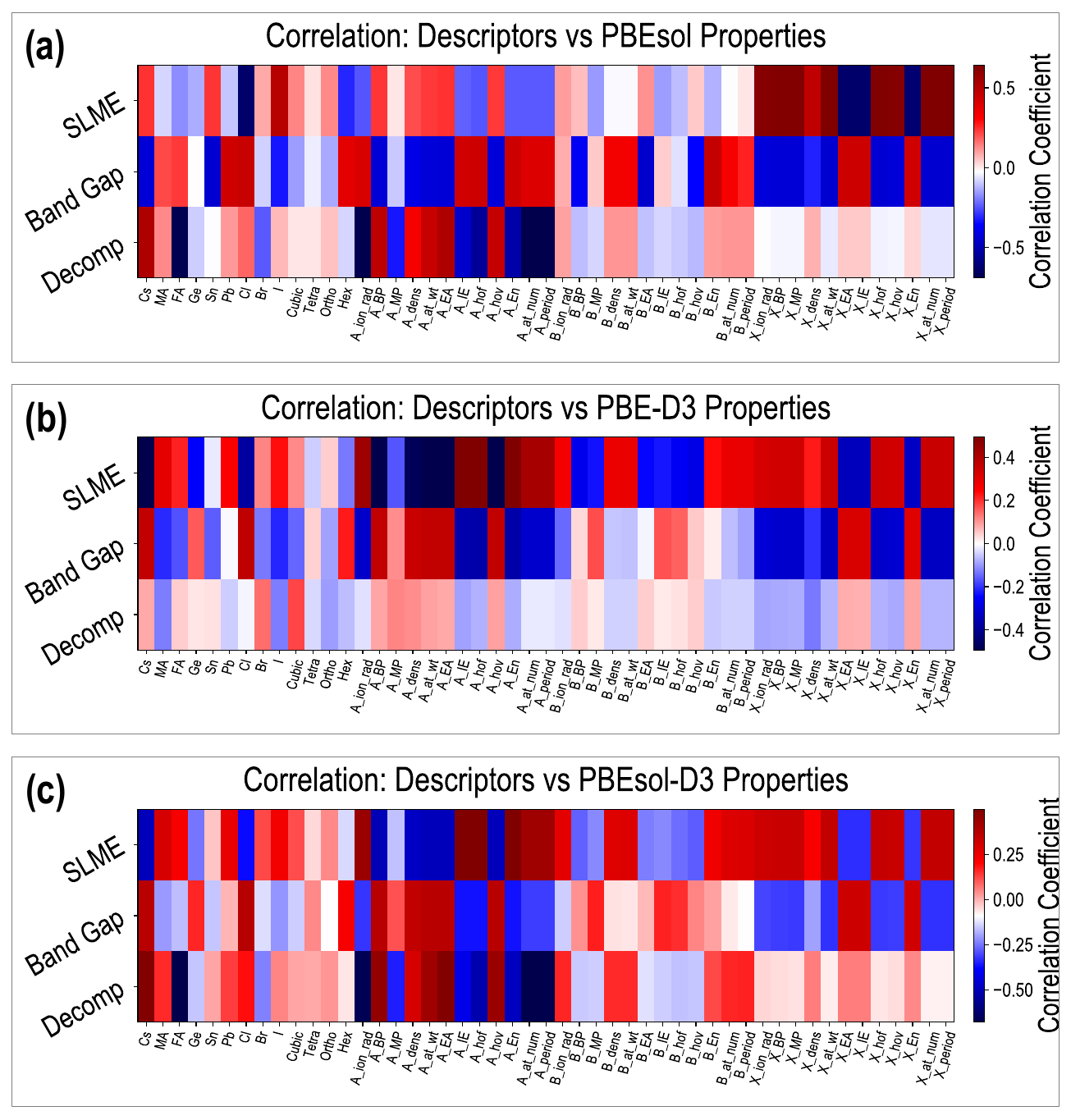}
  \caption{Pearson correlation coefficients between DFT-computed properties ($\Delta$H, E$_{g}$, and SLME) and 49-dimensional input descriptors, for (a) PBEsol, (b) PBE-D3, and (b) PBEsol-D3 datasets of 146 compounds.} \label{fig:S10}
  \end{center}
\end{figure}

\end{document}